\documentclass[a4paper,11pt]{article}
\pdfoutput=1 % if your are submitting a pdflatex (i.e. if you have
\usepackage{jcappub} % for details on the use of the package, please
\usepackage[T1]{fontenc} % if needed

\usepackage{amsmath}
\usepackage{color}
\usepackage{graphicx}
\usepackage{mathtools}
\usepackage[normalem]{ulem}

\usepackage{custom_commands}

\def\gtorder{\mathrel{\raise.3ex\hbox{$>$}\mkern-14mu
             \lower0.6ex\hbox{$\sim$}}}
\def\ltorder{\mathrel{\raise.3ex\hbox{$<$}\mkern-14mu
             \lower0.6ex\hbox{$\sim$}}}

\definecolor{mygreen}{RGB}{0, 100, 0}

\title{Bandpass mismatch error
for satellite CMB experiments I:
Estimating the spurious signal}

\author[a,b]{Duc Thuong Hoang,}
\author[a]{Guillaume Patanchon,}
\author[a,c]{Martin Bucher,}
\author[d,e]{Tomotake Matsumura,}
\author[a]{Ranajoy Banerji,}
\author[f]{Hirokazu Ishino,}
\author[g,e,d,h]{Masashi Hazumi,}
\author[a,i]{Jacques Delabrouille}

\affiliation[a]{Laboratoire Astroparticule et Cosmologie (APC), Universit\'e Paris Diderot, CNRS/IN2P3, 
CEA/Irfu, Observatoire de Paris,
Sorbonne Paris Cit\'e, 10, rue Alice Domon et L\'eonie Duquet, 75205 Paris Cedex 13, France}

\affiliation[b]{Department of Space and Aeronautics, University of Science and Technology of Hanoi (USTH), Vietnam Academy of Science and Technology, 18 Hoang Quoc Viet, Cau Giay District, Hanoi, Vietnam}

\affiliation[c]{Astrophysics and Cosmology Research Unit, School of Mathematics, Statistics and Computer Science, 
University of KwaZulu-Natal, Durban 4041, South Africa}

\affiliation[d]{Kavli Institute for the Physics and Mathematics of the Universe (Kavli IPMU, WPI), UTIAS, The University of Tokyo, Kashiwa, Chiba 277-8583, Japan} % Kavli IPMU

\affiliation[e]{Institute of Space and Astronautical Science (ISAS), Japan Aerospace Exploration Agency (JAXA), Sagamihara, Kanagawa 252-0222, Japan} % ISAS/JAXA

\affiliation[f]{Department of Physics, Okayama University, 3-1-1 Tsushimanaka, Kita-ku, Okayama 700-8530, Japan}

\affiliation[g]{High Energy Accelerator Research Organization (KEK), Tsukuba, Ibaraki 305-0801, Japan} % KEK

\affiliation[h]{The Graduate University for Advanced Studies (SOKENDAI), Miura District, Kanagawa 240-0115, Hayama, Japan} % SOKENDAI
\affiliation[i]{D{\'e}partement d'Astrophysique, CEA Saclay DSM/Irfu, 91191 Gif-sur-Yvette, France}

\abstract{
Future Cosmic Microwave Background (CMB) satellite missions aim to use the $B$ mode  
polarization to measure the tensor-to-scalar ratio $r$ with a sensitivity $\sigma_r \ltorder 10^{-3}$. Achieving this goal will not only require 
sufficient detector array sensitivity but also 
unprecedented control of all systematic errors inherent to CMB polarization
measurements. Since polarization measurements derive from 
differences between observations at different times and from different
sensors, detector response mismatches introduce leakages from intensity
to polarization and thus lead to a spurious $B$ mode signal.
Because the expected primordial $B$ mode 
polarization signal is dwarfed by the known unpolarized intensity signal, such leakages 
could contribute substantially to the final error budget for measuring $r.$
Using simulations we estimate the magnitude and angular spectrum of the spurious
$B$ mode signal resulting from bandpass mismatch between different detectors.
It is assumed here that the detectors are calibrated, for example using the CMB dipole,
so that their sensitivity to the primordial CMB signal has been perfectly matched.
Consequently the mismatch in the frequency bandpass shape between detectors
introduces differences in the relative calibration of galactic emission components. 
We simulate this effect using a range of scanning patterns being considered for future satellite missions. 
We find that the spurious contribution to $r$ 
from reionization bump on large angular scales ($\ell < 10$) is 
$\approx 10^{-3}$ assuming large detector arrays and 20 percent of the sky masked.
We show how the amplitude of the leakage depends
on the angular coverage per pixels that results from the scan pattern.
}

\begin{document}

\maketitle
\flushbottom

\section{Introduction}

Measurements of the cosmic microwave background (CMB) 
provide a rich data set for studying cosmology and astrophysics
and for placing stringent constraints on cosmological models.  In
particular, the ESA Planck satellite mission has produced
full sky maps in both temperature and polarization 
at unprecedented sensitivity in nine broad $(\Delta \nu /\nu \approx 0.3)$
microwave frequency bands~\cite{planck2011-1.1}.

Conventional cosmological models predict that the CMB is linearly polarized,
so that the fourth Stokes parameter $V$ vanishes. CMB polarization patterns can be 
decomposed in two components known as the $E$ and $B$ modes, respectively of even and 
odd parity. In linear cosmological perturbation theory, {\it scalar} perturbations 
produce $E$ mode polarization but are unable to produce any $B$ mode polarization at 
linear order. The $E$ mode polarization angular power spectrum can be predicted from a 
model fitted to the measured $T$ anisotropies. The WMAP~\cite{WMAP13} and Planck~\cite{PlanckSpectra16} space missions, 
complemented on smaller angular scales by ACT~\cite{ACTPol17} and SPT~\cite{SPTPolE15}, have already measured the $E$ 
mode polarization power spectrum up to high multipole number $\ell$, even if the 
accuracy of the measurement can still be substantially improved. On the other hand, the 
odd parity (or pseudo-scalar) polarization pattern called the $B$ mode arises either 
from primordial tensor perturbations, or equivalently primordial gravitational waves, 
presumably generated during inflation, or from scalar modes at higher nonlinear order, 
primarily through gravitational lensing. Gravitational lensing $B$ modes dominate over 
primordial $B$ modes on small angular scales. These gravitational lensing $B$ modes 
have already been observed at $\ell\gtorder 100$ by the POLARBEAR~\cite{Polarbear2014},
SPT-Pol~\cite{SPTPolB15} and Bicep2/Keck~\cite{Bicep2KEK15} ground-based experiments. Primordial $B$ modes have not 
been observed yet. Their predicted shape features a `recombination bump' visible at $
\ell \approx 80,$ and a `reionization bump' at $\ell \ltorder 10$. The overall 
amplitude of this primordial B-mode spectrum depends linearly on the value of the 
tensor--to--scalar ratio $r$. The current upper limit is $r<0.07$ at 95\% c.l.~
\cite{Bicep2Planck,Bicep2Kek}. 

After Planck, a number of ground-based and balloon-borne experiments currently either 
taking data or in the planning stage aim to make the first detection of primordial $B$ 
modes. In parallel, the space-borne mission concepts CORE~\citep{Delabrouille17}, 
LiteBIRD~\cite{Hazumi12,Matsumura2016}, and PIXIE~\cite{Pixie16} have been designed to probe $B$ 
modes at higher sensitivities and using frequency bands inaccessible from the ground.  
Constraining physically--motivated inflation models requires sensitivities in the 
tensor-to-scalar ratio of $\sigma_r \ltorder 10^{-3}$, almost two orders of magnitude 
beyond the Planck sensitivity. Furthermore, systematic errors must be controlled so 
that their contribution to the final error budget is subdominant. The calibration 
requirements become correspondingly more stringent, and future experiments will have to 
devise novel calibration procedures to characterize the instrument at a level that 
makes it possible to correct the raw data at sufficient accuracy.

Typically experiments observe in a number of different frequency channels with many 
detectors for each frequency channel. Ideally, all detectors in a single channel should 
have the exact same bandpass function (i.e the response $g(\nu)$ that defines the 
transmission of the system as a function of frequency) in order to construct single 
band maps, which are then analyzed to isolate the primordial cosmological signal. Many 
detectors are necessary in each channel to improve on the sensitivity of current 
observations, which already use detectors that are very nearly at the quantum noise limit.
If however the detectors that are meant to be identical have slightly different
bandpasses, artifacts are introduced into maps that are obtained by combining the 
signals from several detectors. After cross-calibration on the CMB, for instance using 
the bright CMB dipole, the amplitude of other astrophysical components is different in 
the different detectors, and residuals of the differences of integrated intensity 
project onto the reconstructed polarization maps. Such effects have been observed in 
Planck~\cite{PlanckLowEll} and WMAP~\cite{Komatsu14}. In this paper we call these 
artifacts `bandpass mismatch errors'.

Obviously, such errors can be avoided if the observing strategy allows first to make 
polarization maps with each detector independently, hence without bandpass mismatch 
errors, and then to combine these individual detector maps into a global map. This 
however requires observing each sky pixel with enough independent orientations of the 
detector polarizer. This polarization modulation can be achieved either with the use of 
a rotating half-wave plate (HWP), or by rotating the whole instrument so that each 
pixel is observed with an optimized set of detector orientations. However, practical 
considerations may constrain the range of possible polarization orientations, leading 
to a loss of sensitivity after combining single detector polarization maps.

The objective of this paper is to evaluate the level of the bandpass mismatch effect 
for future space missions, and to estimate its possible impact on the final 
determination of the tensor-to-scalar ratio $r$ if no correction measures are taken.  
Our study first focuses on the case without a HWP, and we also verify that the effect 
is greatly reduced with an ideal rotating HWP without any achromaticity or other
non-idealities. For a more detailed discussion of general issues pertaining to the use 
of a HWP for achieving polarization modulation and in particular a discussion of the 
issue of achromaticity, we refer the reader to the results of the ABS
experiment~\cite{ABS16} and the thesis~\cite{TomoThesis} and references therein. We note that in 
the first case, making single detector maps 
that are subsequently combined to avoid band--pass mismatch errors, could be done at the price of increased final noise since the angular 
coverage in each pixel is sub-optimal. HWP non-idealities are not studied in this 
paper.

In Sect.~\ref{SectTwo} we model the bandpass mismatch effect, and in
Sect.~\ref{SectThree} we evaluate the impact on $B$ mode measurements and relate the 
mismatch errors to the "crossing moment maps", that provide a measure of uniformity of 
polarizer angle coverage in each pixel. Correction methods are developed in a companion 
publication~\cite{BPMMII}.

\section{Sky emission model and mismatch errors}
\label{skyModel:sect}

\label{SectTwo}

The total intensity of the microwave sky 
can be expressed as a sum of components of different astrophysical origin. 
In intensity, the CMB anisotropies are dominant over most of the sky, but several diffuse components 
of Galactic origin are also present as well as compact sources, which include
extragalactic radio sources, IR sources (understood to be dusty
galaxies), and Sunyaev-Zeldovich (SZ) distortions from the hot gas within
galaxy clusters. We model the unpolarized sky
emission at position $\hat p $ and frequency $\nu$ as 
\begin{equation}
  I_{\rm tot}(\hat p ,\nu) = I_0(\nu )+ 
{\partial B(\nu ; T) \over \partial T}\Bigg| _{T_0}\,\Delta T_{\rm CMB}(\hat p ) + \sum_{(c)} I_{(c)}(\hat p ,\nu )
\end{equation}
where $B(\nu ; T)$ is the spectrum of a blackbody at temperature $T$, $T_0$ is the average CMB temperature of about 2.7255\,K,
$\Delta T_{\rm CMB}(\hat p )$
is the CMB temperature fluctuation around this mean value, $I_{(c)}(\hat p ,\nu )$ the emission 
spectrum of component $(c)$ as a function of electromagnetic frequency $\nu$,
$I_0(\nu )$ is the monopole including all components. We have 
similar relationships for the $Q$ and $U$ Stokes parameters. All three Stokes parameters of the CMB at a given position on
the celestial sphere have the factorized frequency dependence as given 
above. A similar factorizable form can be used for 
the SZ emission assuming that the hot gas is non-relativistic.
The galactic components are more complicated at the accuracy required for future
satellite missions and an Ansatz where the frequency dependence 
of each component factorizes out breaks down. However for studying
bandpass mismatch error to first order, a simple factorizable model suffices. 

For this bandpass mismatch study, we consider only the CMB and the diffuse 
galactic components, which contribute the largest bandpass mismatch effects.  
At frequencies $\approx 150$\,GHz where
we focus our study, the galactic emission can be
decomposed into thermal dust emission, which is the dominant component,
and synchrotron, free-free, and spinning dust emissions. The carbon
monoxide (CO) rotational emission at transition line frequencies
$\nu=115$\,GHz for $J = 1 \rightarrow 0$ and $\nu=230$\,GHz for $J = 2
\rightarrow 1$ was a source of significant leakage
in Planck experiment~\cite{PlanckCO2013}, but is not considered here
because we anticipate that
the filters used by future satellite
experiments will avoid these lines.

For our study we assume that the galactic thermal dust emission
is a greybody of temperature $T_d \approx 19.7$\,K
\citep{PlanckDust13} with an emissivity spectral index $\beta(\hat p ),$ 
which depends on sky position and whose average value is $\approx 1.62$ 
as measured by Planck
\cite{PlanckDust13,PlanckDustPolar14}. The synchrotron and free-free emissions
can be described by power law spectra with the negative spectral indices $\approx
-3.1$ and $\approx -2.3$, respectively (see \cite{PlanckCompSep15} and
references therein). The fluctuation of the signal (relatively to the average CMB monopole) 
measured by the detector $i$ is given by
\begin{multline}
\int d\nu ~g_i(\nu)~\Bigl(I(\hat p ,\nu ) - I_0(\nu )\Bigr)
= \int d\nu ~g_i(\nu) {\partial B(\nu ;T) \over \partial T}\Big| _{T_0} 
\Delta T_{\rm CMB}(\hat p )\\ 
+ \int d\nu ~ g_i(\nu)~I_d(\hat p , \nu _0)~ 
\left({\nu\over \nu_0}\right)^{\beta(\hat p )} 
\dfrac{B(\nu ; T_d)}{B(\nu _0; T_d)}
+ \ldots,
\label{eq:intfilt}
\end{multline}
where $I_0(\nu )=B(\nu ; T_0)$ is the CMB monopole, $g_i(\nu)$ is the bandpass function 
of the detector $i$, $I_d(\hat p, \nu_0)$ is the amplitude of the dust component at the 
reference frequency $\nu _0$, and where the dots stand for other components (such as 
synchrotron and free-free) not explicitly written here. To first order we obtain for 
the total sky intensity $I_{\rm sky}(\nu _0)$ after converting the CMB temperature $\Delta T_{\rm CMB}$ to intensity $I_{\rm CMB}(\nu _0)$:
%\begin{equation}
%\Delta T_{\rm sky} = 
%\Delta T_{\rm CMB} + 
%\gamma_d\,\Delta T_{\rm dust}(\nu _0) 
%+ \gamma_s\,\Delta T_{\rm sync}(\nu _0) + \ldots 
%\label{eq:sumcompT}
%\end{equation}
\begin{equation}
I_{\rm sky}(\nu _0) = 
I_{\rm CMB}(\nu _0) + 
\gamma_d\,I_{\rm dust}(\nu _0) 
+ \gamma_s\,I_{\rm sync}(\nu _0) + \ldots,
\label{eq:sumcomp}
\end{equation}
where
\begin{equation}
\gamma_d = 
\left(
\frac{\int d\nu ~g_{i}(\nu) 
\left(\frac{\nu}{\nu_{0}}\right)^\beta 	           
\frac{B(\nu ;T_{d})}{B(\nu_{0}; T_{d})}}  
{\int d\nu g_{i}(\nu) \left( \frac{\partial{B(\nu ; T)}}{\partial{T}} \right) \Big| _{T_{0}}}
\right) 
\left( \frac{\partial{B(\nu_{0}; T)}}{\partial{T}} \right) \Big| _{T_{0}}.
\label{eq:intfilters}
\end{equation}
The factor $\gamma_s$ is similarly defined integrating over the synchrotron spectrum, etc. 
%From eqn.~(\ref{eq:sumcompT}) we convert each term and write the same relationship for the intensity
%\begin{equation}
%I_{\rm sky}(\nu _0) = 
%I_{\rm CMB}(\nu _0) + 
%\gamma_d\,I_{\rm dust}(\nu _0) 
%+ \gamma_s\,I_{\rm sync}(\nu _0) + \ldots
%\label{eq:sumcomp}
%\end{equation}
%where we drop the $\Delta$ even if the quantities are variations with respect to the monopole.

Eqn.~(\ref{eq:sumcomp}) also holds for the polarization when $I$ is replaced 
with $Q$ and $U.$
The unit normalization for the CMB component 
is justified because the data are calibrated using the CMB dipole (or higher order temperature anisotropies). The
values of the $\gamma$ parameters are close to unity when the bandwidth
is narrow. %For typical bandwidths of CMB experiments and $\nu_0$ 
%chosen near the band center, and $\gamma_s$ and $\gamma_d$ have means away
%from one and also some variance about this mean owing to bandpass mismatch.

Differences in the bandpass function $g_i(\nu)$ from detector to detector
result in corresponding 
variations in $\gamma $ from detector to detector
for each non-CMB component. Such variations have
been observed in Planck data (see Figs.~5 and~28 of~\cite{PlanckIX2013} for 
the measured Planck filters and the mismatch parameters, respectively).
Pre-flight Fourier Transform Spectrometer (FTS) ground measurements characterized
variations of the filter edge positions 
at both the low and high frequencies at about the percent level.  
Ground measurements, however, were not accurate enough to detect 
variations near the center of the filters, and thus 
could not be used to determine the 
$\gamma $ parameters with sufficient accuracy. The $\gamma $ parameter
variations had to be determined from flight data  to allow an accurate correction of the
leakage (see the low-$\ell $
Planck paper \cite{PlanckLowEll} as well as~\cite{PlanckCalib2015}). 
It should be noted that the variations 
of the bandpass functions of the filters from detector to detector for a future satellite
experiment will depend on the kind of detector technology used (see also~\cite{Jarosik07} regarding the WMAP experiment).

As already stressed, for the above sky emission model where each component has a 
fixed (factorizable) frequency dependence, the bandpass mismatch
maps depend only on the $\gamma$ parameters and not on the other 
details of the filters. The deviations from this simplified model due 
to the observed spatial variations of the spectral indices of component spectra 
and of thermal dust temperature produce a second order correction to the bandpass mismatch effect, which is neglected for this study.
Consequently, the intensity to polarization leakage due to
bandpass mismatch can be obtained using only the $\gamma $'s and no
additional properties of the bandpass functions.

To relate these variations to filter properties, 
we assume a simple model in which each frequency band is a tophat bandpass function for which $g(\nu)=1$ in the interval $[\nu _{\rm min}, \nu _{\rm max}]$ 
and $g(\nu)=0$ elsewhere. We 
assume that the variations in $\nu _{\rm min}$ and $\nu _{\rm max}$ for each detector 
are generated independently according to a uniform distribution with a width of 1\%.\footnote{
We thank Aritoki Suzuki for sharing with us that the measurement errors with FTS in the 
bandpass of the third-order Chebyshev filter placed between the 
broadband sinuous antennas and the bolometers of the focal plane 
panels of the Simons Array \cite{Westbrook} give approximately this spread. 
Obviously, since these are values dominated by measurement error, the actual
bandpass mismatch for these filters could be much smaller.
These measurements merely serve to establish an upper bound on the mismatch.
These values are also of the same order of magnitude as the 
values representing the bandpass mismatch of the metal mesh filters used 
as part of the Planck satellite HFI instrument. [See \cite{2016A&A...596A.107P} 
for a discussion of the Planck bandpass mismatch.]}
%\begin{equation}
%{\rm rms}\bigg[{\Delta \nu _{\rm min} \over \nu_0}\bigg]=
%{\rm rms}\bigg[{\Delta \nu _{\rm max} \over \nu_0}\bigg]=1\% .
%\end{equation}

We also assume a bandwidth $(\nu_{\rm max}-\nu_{\rm min})/\nu_0$ of 0.25 on average, with $\nu_0=140.7\,$GHz.  
The resulting 
RMS of $\gamma_d$ is 0.6\% for this simple model. 
This is similar to the variations observed for Planck at 143\,GHz.
The fact that actual bandpass functions are 
more complex functions of $\nu$ does not affect the applicability of the present work as
long as the corresponding $\gamma$ coefficients remain of the same order of magnitude. 
Results for other values may be obtained by trivial rescaling.
We verified the expected linear scaling 
by increasing the width of the uniform distribution
from 1\% to 2\% and observed that the leakage increases by
a factor of 2, as expected.

This simple model for detector bandpasses is appropriate for the
foreground components having a smooth frequency spectral dependence
(e.g., synchrotron and dust emission), but for galactic line emission (such
as galactic CO emission and other spectral lines) a more detailed model would be required. 
The $\gamma $'s are computed as a random set from this distribution model, 
since those are the only quantities needed for the bandpass mismatch evaluation. 

In this paper, we focus our analysis on a frequency channel centered 
at $\nu_0=140$\,GHz, and so we restrict ourselves to the dominant galactic component, namely the thermal 
dust emission. More galactic components are included in the companion paper 
discussing the correction of the mismatch~\cite{BPMMII}.

\section{Calculating the bandpass mismatch}

\label{SectThree}

In this Section we use a simplified model of the measurement, stripped of additional 
complications such as asymmetric beams, pixelization effects, etc.~for estimating the 
dominant contribution to the bandpass mismatch error. A study of more than one source 
of systematic errors simultaneously would obviously be more complicated and also less 
intuitive to interpret. Here our purpose is to study bandpass mismatch error in 
isolation and in the simplest possible context.

\begin{figure}[h]
\begin{center}
\includegraphics[width=0.7\textwidth ]{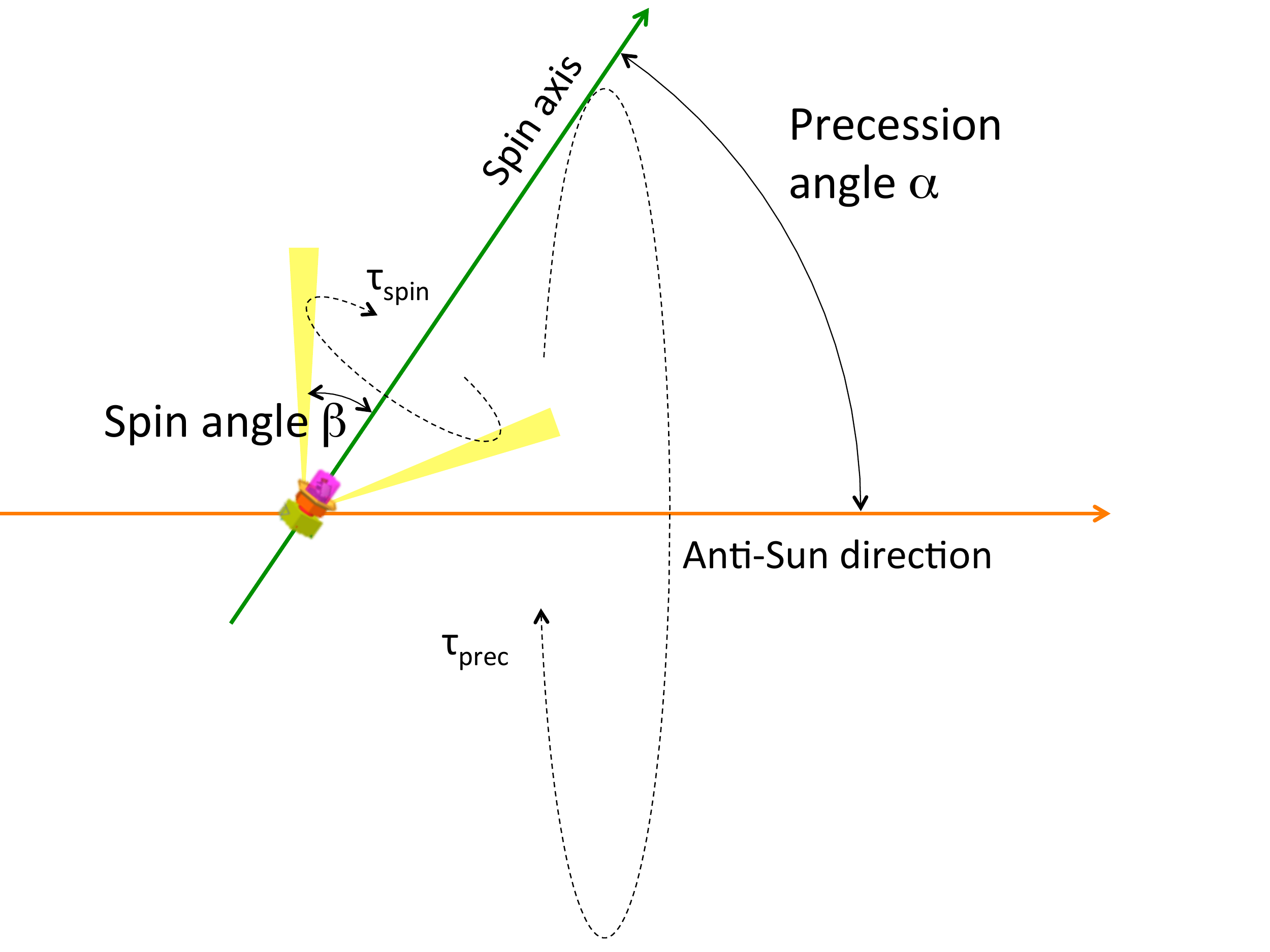}
\end{center}
\caption{Representation of typical satellite scanning strategy.}
\label{fig:scan}
\end{figure}

We assume a scanning pattern that combines three rotations: a relatively fast spin of 
the payload, around a spin axis that precesses around the anti-solar direction, which 
itself follows the yearly motion of the spacecraft around the Sun. Many of the proposed 
future CMB polarization space missions have adopted such a scan strategy 
\cite{EPIC2009, Matsumura2016, Delabrouille17}. The exact scanning pattern is 
characterized by the following parameters: $\alpha $ (precession angular radius), $
\beta$ (spin angular radius), $\tau _{\rm prec}$ (precession period), and
$\tau _{\rm spin}$ (spin period). The motion of the satellites and the definitions of 
the scanning parameters are indicated in Figure~\ref{fig:scan}.

Our simulations use maps of the celestial sphere 
pixelized using HEALPix\footnote{http://healpix.sourceforge.net}~\cite{Healpix05} (with $n_{\rm side}=256$). A sufficiently fast
sampling rate is chosen so that several hits are recorded during each
pixel crossing, so that altering this parameter does not significantly affect 
the results. White instrument noise of a stationary amplitude is assumed, and 
under this hypothesis, we solve the map making equation:
\begin{equation}
\widehat {\mathbf{m}}
=
(\mathbf{A}^T \mathbf{N}^{-1} \mathbf{A})^{-1} 
(\mathbf{A}^T \mathbf{N}^{-1} \mathbf{d}).
\label{mapMakingEquation}
\end{equation}
Here the notation is such that $\widehat {\mathbf{m}}$ includes the estimated maps of 
Stokes parameters $\widehat{I},$ $\widehat{Q}$ and $\widehat{U}$; $\mathbf{A}$ is the 
pointing matrix relating data samples to map; $\mathbf{N}$ is the noise covariance 
matrix in the time domain; and we denote the  polarization angle of a detector $\psi $ 
with respect to a reference axis. Individual measurements comprising the data vector
$\mathbf{d}$ are given by
\begin{equation}
S_j=
I(p)+
Q(p)\cos 2\psi _j+
U(p)\sin 2\psi _j
+n _j
\label{eqSIQU}
\end{equation}
where $n _j$ represents a stationary white noise source for observations indexed by $j.$ Here the index $j$ ($j=1,\ldots ,N_p$) 
labels the observations falling into the pixel labelled by $p.$
The normalization of the noise does not matter for our purpose.
The model here assumes that all the beams are azimuthally symmetric and identical.

The hypothesis of white instrument noise provides considerable simplification because 
in this special case the map making equation [i.e., eqn.~(\ref{mapMakingEquation})]
can be cast into a block diagonal form, so that 
the equations for different pixels decouple from each other. Each block 
(labelled by the pixel index $p$) takes the form
\begin{eqnarray}
\begin{pmatrix}
\phantom{\bigg|}\widehat  I(p)\\
\phantom{\bigg|}\widehat Q(p)\\
\phantom{\bigg|}\widehat U(p)
\end{pmatrix}
&=&
\dfrac{1}{N_p}\times
\begin{pmatrix}
\phantom{\bigg|}1 & 
\phantom{\bigg|}\left< \cos 2\psi_j \right> &
\phantom{\bigg|}\left< \sin 2\psi_j \right> \\
\phantom{\bigg|}\left< \cos 2\psi_j \right> &
\phantom{\bigg|}\dfrac{1+\left< \cos 4\psi_j \right> }{2} &
\phantom{\bigg|}\dfrac{\left< \sin 4\psi_j \right> }{2} \\
\phantom{\bigg|}\left< \sin 2\psi_j \right> &
\phantom{\bigg|} \dfrac{ \left< \sin 4\psi_j \right> }{2}  &
\phantom{\bigg|} \dfrac{1- \left< \cos 4\psi_j \right> }{2}
\end{pmatrix}^{-1}\cr
&&\times 
\begin{pmatrix}
\phantom{\bigg|}\sum _j S_j~\phantom{\cos 2\psi _j}\\
\phantom{\bigg|}\sum _j S_j~\cos 2\psi _j\\ 
\phantom{\bigg|}\sum _j S_j~\sin 2\psi _j
\end{pmatrix}
\label{myEquationThree}
\end{eqnarray}
where the hats indicate the maximum likelihood estimator,
and $\left< . \right>$ denotes the average of a quantity
over all data samples $j$.

Computing the maps $\widehat I(p),$ $\widehat Q(p),$ and $\widehat U(p)$
as above gives the minimum variance estimator of the 
sky signal in the frequency band under consideration under
the hypothesis that the noise of each detector is white
(with no correlation in time giving rise to excess low-frequency 
noise, nor variation of the noise r.m.s. with time),
that it is uncorrelated between detectors, and that its level 
is identical in all detectors \cite{Couchot99}. 
It also assumes no source of systematic errors which may require
a different detector weighting to estimate each of the Stokes parameters
(and, in particular, no bandpass mismatch).

Following Equation~\ref{eqSIQU}, for this map-making solution, bandpass mismatch causes 
the following map errors
\begin{eqnarray}
\begin{pmatrix}
\phantom{\bigg|}\delta \widehat I_{BPM}\\
\phantom{\bigg|}\delta \widehat Q_{BPM}\\
\phantom{\bigg|}\delta \widehat U_{BPM}
\end{pmatrix}
&=&
\begin{pmatrix}
\phantom{\bigg|}1 & 
\phantom{\bigg|}\left< \cos 2\psi_j \right>  &
\phantom{\bigg|}\left< \sin 2\psi_j \right>  \\
\phantom{\bigg|}\left< \cos 2\psi_j \right>  &
\phantom{\bigg|}\dfrac{1+
\left<
\cos 4\psi_j
\right>
}{2} &
\phantom{\bigg|}\dfrac{
\left<
\sin 4\psi_j
\right>
}{2} \\
\phantom{\bigg|}
\left<
\sin 2\psi_j
\right>
&
\phantom{\bigg|}\dfrac{\left< \sin 4\psi_j \right> }{2} &
\phantom{\bigg|}\dfrac{1-
\left<
\cos 4\psi_j
\right>
}{2}
\end{pmatrix}^{-1} \cr
&&\times 
\begin{pmatrix}
\phantom{\bigg|}\delta \left< S_j\right>\\
\phantom{\bigg|}\delta \left< S_j\cos 2\psi _j\right>\\ 
\phantom{\bigg|}\delta \left< S_j\sin 2\psi _j\right>
\end{pmatrix}
\label{bpmErrorMatEqn}
\end{eqnarray}
where $\delta \left< S_j\right>,$  $\delta \left< S_j\cos 2\psi _j\right>,$ 
and $\delta \left< S_j\sin 2\psi _j\right>$
are functions of the underlying sky component maps.
Here we assume that the normalization of the CMB
component for each detector is perfect. This is obviously an idealization
because in reality there are also systematic errors from uncorrected
gain variation, but this is a separate issue that we do not analyze 
here. Moreover, since the relative gain of the detectors is
calibrated using the CMB dipole, the approximation that the error is mostly in
the relative contributions of the other components is a reasonable one.

Given a model 
of the microwave sky, the bandpass functions of the various detectors, and the scanning pattern on the sky,
eqn.~(\ref{bpmErrorMatEqn}) can be used to compute the bandpass mismatch errors in the reconstruction
of a map of Stokes parameters.
For future studies of the CMB polarization, and in particular for the search
for primordial $B$ modes, the error of greatest concern arises from the leakage
of the $I$ component of the foregrounds into the $Q$ and $U$ components of the 
maximum likelihood band sky maps.
From eqn.~(\ref{bpmErrorMatEqn}) we observe that the 
three terms: $\delta \left< S_j\right>,$  $\delta \left< S_j\cos 2\psi _j\right>,$ 
and $\delta \left< S_j\sin 2\psi _j\right>$ can potentially induce a bias on the 
polarization Stokes parameters. The first term $\delta \left< S_j\right>$ has no impact if the maps of $\left< \cos 2\psi \right> $ and 
$\left< \sin 2\psi \right> $ vanish. 
This is the case in particular if detectors are arranged in sets of perfectly orthogonal 
pairs observing the sky along the same scanning path. If in addition for each such pair there is a 
matching pair observing at an angle of $45^\circ$ relative to the first one, we get an optimized configuration
\cite{Couchot99}
for which the 3$\times$3 matrix in eqn.~(\ref{myEquationThree}) takes the form
\begin{equation}
\begin{pmatrix}
1 & 0 & 0\\
0 & \dfrac{1}{2} & 0 \\
0 & 0 & \dfrac{1}{2}
\end{pmatrix} ^{-1}.
\end{equation}
This simple form is preserved when observations are made with a set of such `optimized configurations'
oriented at any angle with respect to each other.
This type of detector arrangement was used for the Planck mission, and is now standard for all proposed CMB polarization experiments. We then get
\begin{eqnarray}
\delta \widehat Q_{BPM}(p)&=&{2}\delta \left< S_j\cos 2\psi _j\right>,\cr
\delta \widehat U_{BPM}(p)&=&{2}\delta \left< S_j\sin 2\psi _j\right>,
\end{eqnarray}
where under the sky model presented in Sect.~\ref{skyModel:sect}
\begin{eqnarray}
\delta \left< S_j\cos 2\psi _j\right>&=&
\sum _{(c)} I_{(c)}(p) \sum _i \gamma _{(c),i} f_i(p) \left< \cos 2\psi _{i,j}\right>  ,\cr 
 \delta \left< S_j\sin 2\psi _j\right> &=&
\sum _{(c)} I_{(c)}(p) \sum _i \gamma _{(c),i} f_i(p) \left< \sin 2\psi _{i,j}\right>  .
\label{eq:leakModGen}
\end{eqnarray}
Here the index $(c)$ labels the non-CMB components of the sky model and $i$
labels the detectors of the frequency channel under consideration (ideally supposed to have the same bandpass
function). The coefficients $\gamma _{(c),i}$ vary from detector to detector as a 
function of the stochastic realizations for the bandpass variation 
$\delta \nu _{\rm min,i}$ and $\delta \nu _{\rm max,i}.$
$f_i(p)$ denotes the fraction of the total hits in pixel $p$ from the detector $i,$ and
$\left< \cos 2\psi _{i,j}\right> $
and 
$\left< \sin 2\psi _{i,j}\right> $ 
are the components of the second-order crossing moments in pixel $p$ for the detector $i.$

\begin{figure}[h]
\begin{center}
\includegraphics[width=0.49\textwidth ]{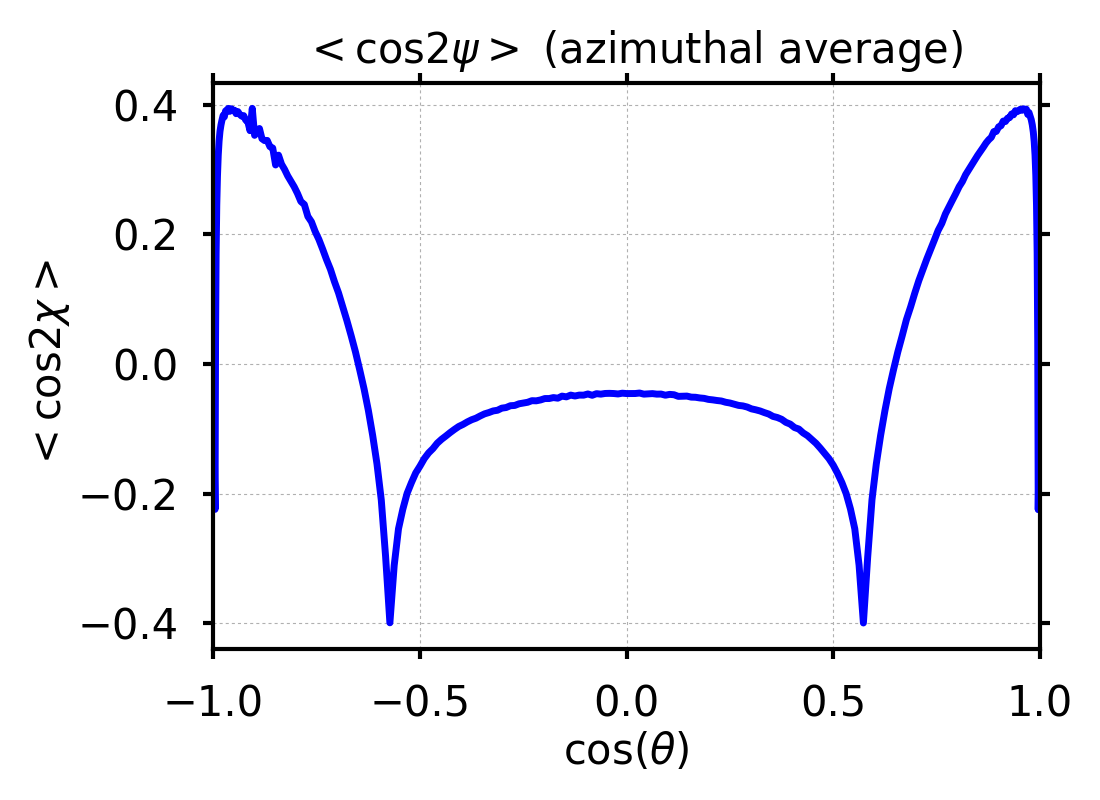}
\includegraphics[width=0.49\textwidth ]{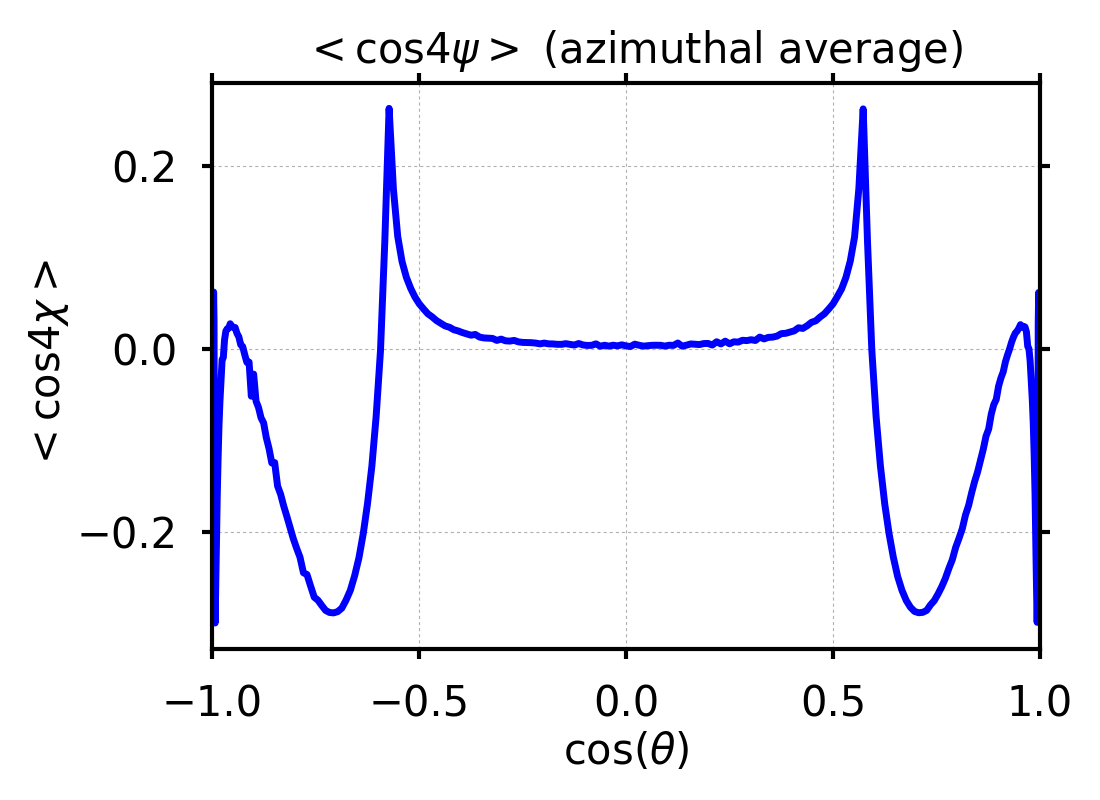}
\end{center}
\caption{{\bf Crossing moment map azimuthal averages.} We show the azimuthal average of 
$\left< \cos 2\psi \right> $ and 
$\left< \cos 4\psi \right> $ 
maps, constituting the totality
of the component that is coherent on large angular scales.
The corresponding
$\left< \sin 2\psi \right> $ and
$\left< \sin 4\psi \right> $ maps vanish for symmetry reasons.}
\label{Fig:CrossLinkingAzimuthalAverages}
\end{figure}

Before describing the predictions of the level of residual due to bandpass mismatch, 
we briefly digress to examine the properties of the crossing moment maps
$\left< \cos 2\psi \right> $, $\left<\cos 4\psi \right> $, $\left< \sin 2\psi \right> $ 
and $\left< \sin 4\psi \right> $ for an individual detector for our model scanning 
pattern characterized by the
parameter values: $\alpha =65^\circ ,$ $\beta =30^\circ ,$ $\tau _{\rm spin}=10.002$ min,
and $\tau _{\rm prec}=96.2079$ min. Those maps, which are studied into more detail
in Sect.~\ref{sub:resonances}, enter into the expression of the
bandpass mismatch. In ecliptic coordinates, 
these quantities have a nearly symmetric pattern around the poles.
Figure~\ref{Fig:CrossLinkingAzimuthalAverages} shows the azimuthally averaged quantities (i.e., 
averaged over the ecliptic angle $\phi$ or ecliptic longitude) as a function 
of the sine of the latitude of the maps.
We observe that for a large fraction of pixels the spin-2 and spin-4 quantities (functions of period $\pi$ and $\pi/2$, respectively) are less then 0.2.

% ---FIGURE---  %%%%%%%%%%%%%%%%%%%%%%%%%%%%%%
%%%%%%%%%%%%%%%%%%%%%%%%%%%%%%%%%%%%%%%

\subsection{Results}

% ---FIGURE---  %%%%%%%%%%%%%%%%%%%%%%%%%%%%%%
\begin{figure}[h]
\centering
\includegraphics[width=0.49\textwidth]{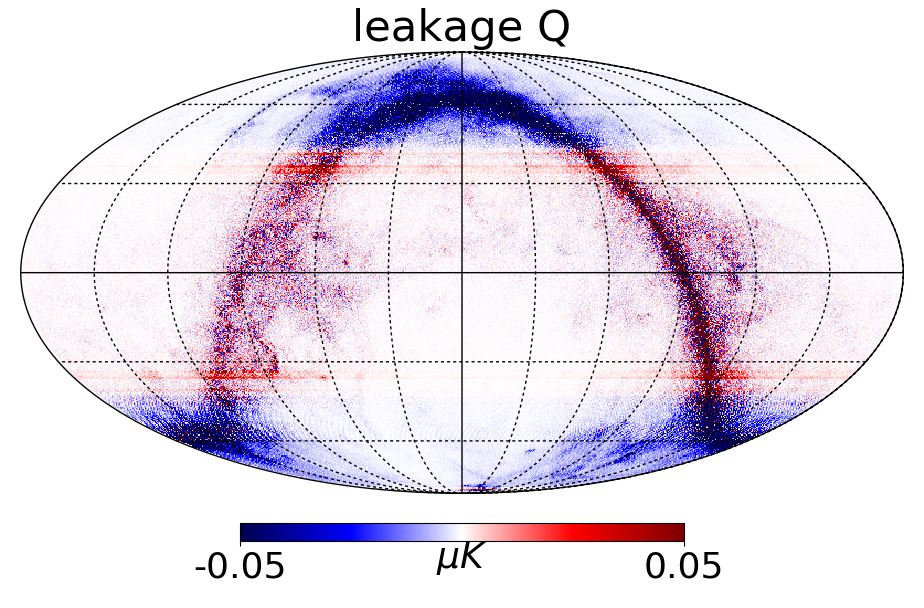}
\includegraphics[width=0.49\textwidth]{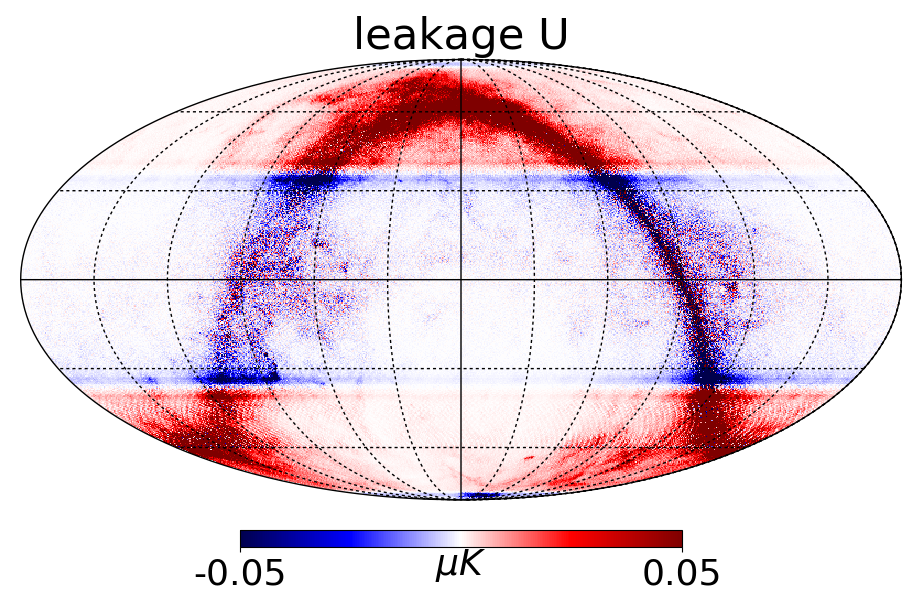}
\caption{$Q$ and $U$ leakage maps, in ecliptic coordinates, with fiducial scanning 
parameters and $N_{\rm det}=222$.}
\label{MapLeak}
\end{figure}
%%%%%%%%%%%%%%%%%%%%%%%%%%%%%%%%%%%%%%%

We now present numerical results for the bandpass mismatch maps and their power spectra 
based on simulations. We construct timestreams for each detector by reading a CMB map 
and a Galactic map, both at $n_{\rm side}=256$, which were preconvolved with a 
symmetric Gaussian beam of $\theta_{\rm FWHM}=32'$. We use an instrument model with 
actual locations of detectors in the focal plane as described 
in~\cite{Matsumura2016} or~\cite{Delabrouille17}. 
depending on the case being considered. We note however that the details of the 
arrangement of the detectors have little or no impact on the leakage power spectra. The 
Galactic map is rescaled from detector to detector using random errors in the bandpass 
generated as described in detail in Sect.~\ref{skyModel:sect}. Then we construct 
combined $I,$ $Q,$ and $U$ maps obtained by applying the map making equation as given 
in eqn.~(\ref{mapMakingEquation}). No noise is included in the simulation, because the 
map making method is linear and the noise does not affect the bias induced by the 
mismatch. For the same reason we do not introduce sky emission polarization in 
simulations. The bandpass mismatch properties of each detector are generated randomly 
and in a statistically independent manner. Figure~\ref{MapLeak} shows the $Q$ and $U$ 
leakage maps: $\delta Q_{\rm BPM}$ and $\delta U_{\rm BPM}$ for one particular 
realization. The output polarization maps result from optimal map making using our 
simulated noiseless and polarizationless timestreams for the\,140\,GHz channel and are 
shown in ecliptic coordinates. The simulation assumed 222 detectors, which is the 
number of detectors composing the LiteBIRD arrays described in~\cite{Matsumura2016}, 
spread over a large focal plane approximatively 10 degrees wide observing with no HWP.
The detector polarizer covers the full range of angles in the focal plane with 
22.5 degree separation. We assume the fiducial scanning parameters 
$\alpha=65^\circ$, $\beta=30^\circ,$ $\tau _{\rm spin}=10$\,min, and $\tau _{\rm prec}= 96.1803$\,min
for the center of the focal plane (see Section~\ref{sub:resonances} for a discussion of 
the choice of $\tau_{\rm spin}$ and $\tau_{\rm prec}$ to minimize the inhomogeneity of 
the scanning pattern which is responsible for Moir\'e effects in the crossing moment 
maps). At 140 GHz the bandpass mismatch error in polarization is dominated by the $I$ 
component of the thermal dust emission, although there are subdominant contributions 
from the diffuse Galactic synchrotron emission and other non-primordial (non-CMB) 
components. The length of the survey in this simulation is exactly one sidereal year in 
order to ensure as uniform and complete sky coverage as possible and hence facilitate 
the interpretation of those results. We observe that the leakage is concentrated near 
the Galactic plane, as expected. The bands at equal latitude visible 
in the leakage maps correspond to regions where the second order crossing moments 
depart significantly from zero (Fig.~\ref{Fig:CrossLinkingAzimuthalAverages}), and as can be seen from eqn.~(\ref{eq:leakModGen}),
there is a strong correlation between the relative leakage amplitude and these moments.

\begin{figure}
  \centering
  \includegraphics[width=1\columnwidth]{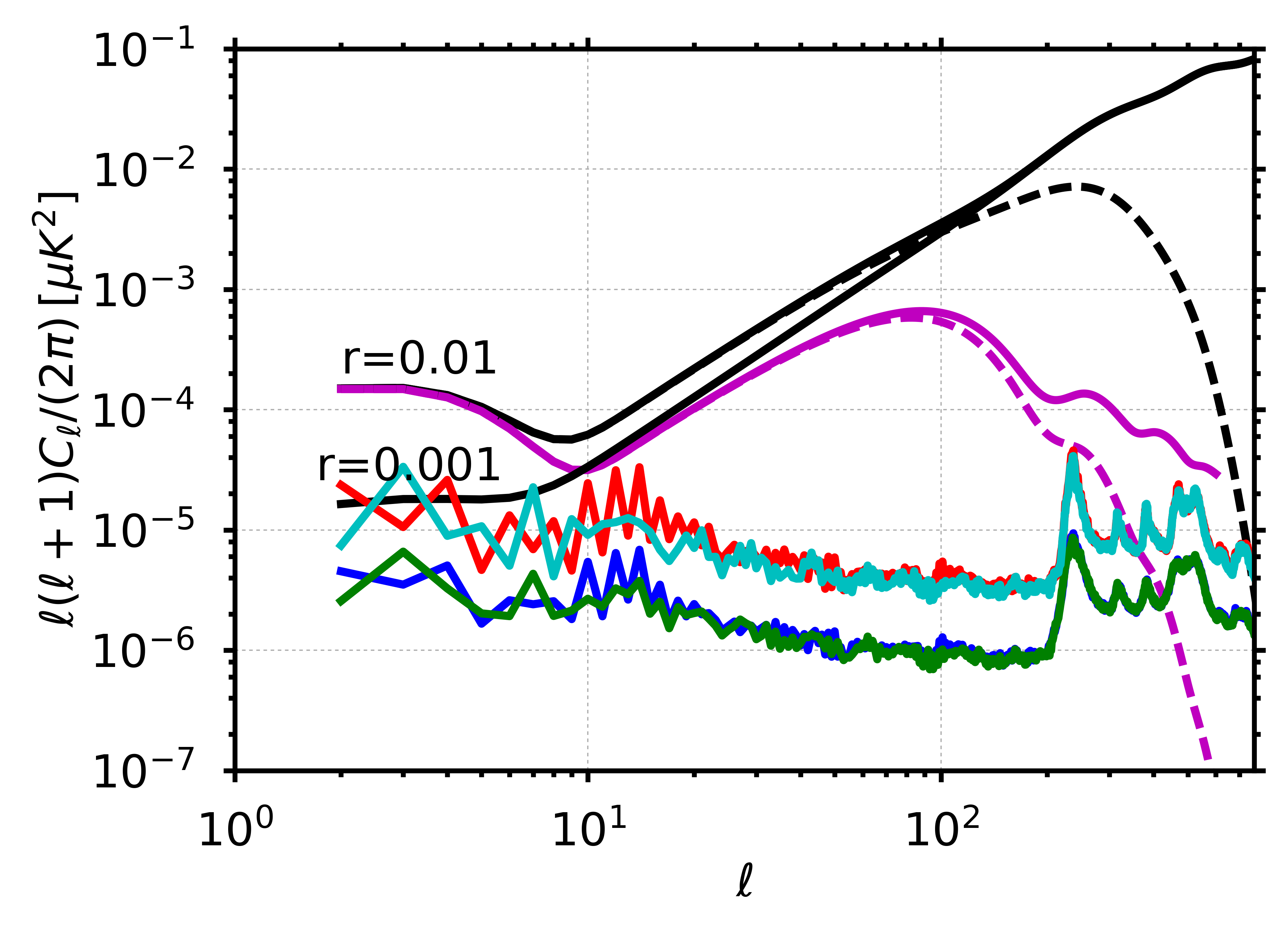}
  \caption{$EE$ and $BB$ leakage power spectra
    for $\alpha=65^\circ$, $\beta=30^\circ$, $\tau_{\rm spin}$ = 10\,min, $\tau_{\rm prec}$= 96.1803\,min,
     and combining data for either 74 or 222 detectors. The red curve corresponds to $BB$
      with 74 detectors, the cyan to $EE$ with 74 detectors, the blue to $BB$ with 222 
      detectors and the green to $EE$ with 222 detectors. The purple curve represents a 
      model of primordial $B$ mode power spectrum with fiducial cosmological parameters 
      after Planck for $r=0.01$, the black curves are including lensing for $r=0.01$ and $r=0.001$. The dashed curves show the effect
      of convolving with a 32 arcmin beam. This plot demonstrates the $1/N_{\rm det}$ 
      dependance of the level of the power spectra.}
    \label{SpectrLeak1overN}
\end{figure}

\begin{figure}[ht]
  \centering
  \includegraphics[width=1\textwidth]{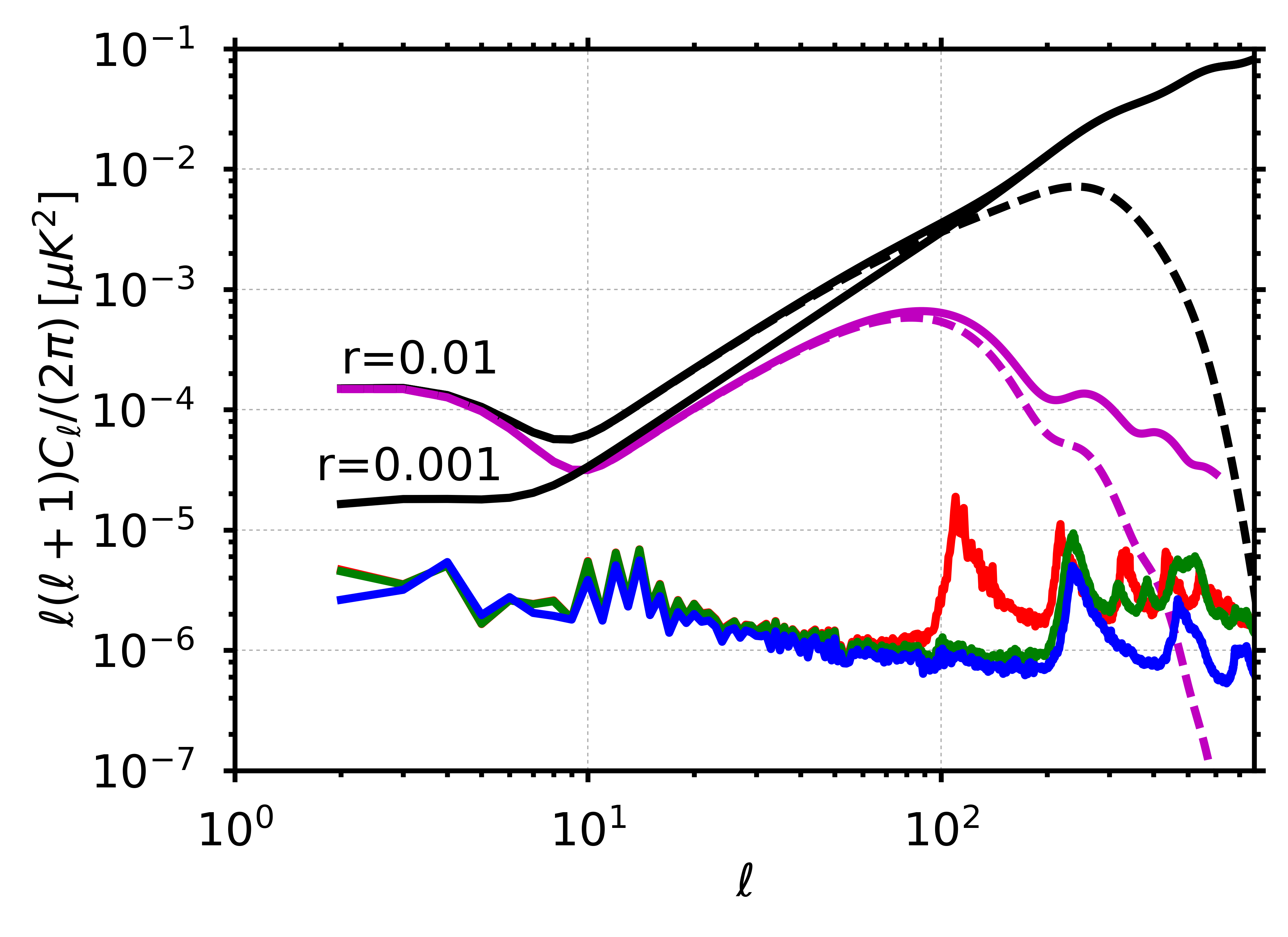}
  \caption{$BB$ leakage power spectra
    for $\alpha=65^\circ$, $\beta=30^\circ$, $\tau_{\rm spin}$=10\,min, $\tau_{\rm prec}$=93\,min (red); $\tau_{\rm spin}$=10\,min, $\tau_{\rm prec}$=96.1803\,min (green); and $\tau_{\rm spin}$=10/3\,min, 
    $\tau_{\rm prec}$=96.1803\,min (blue). 
    Simulations include 222 detectors and 365 days observation. See the Fig.~\ref{SpectrLeak1overN} 
    caption for a description of the model curves.}
    \label{SpectrLeak0.3-0.1}
\end{figure}

\begin{figure}
\centering
\includegraphics[width=1\columnwidth]{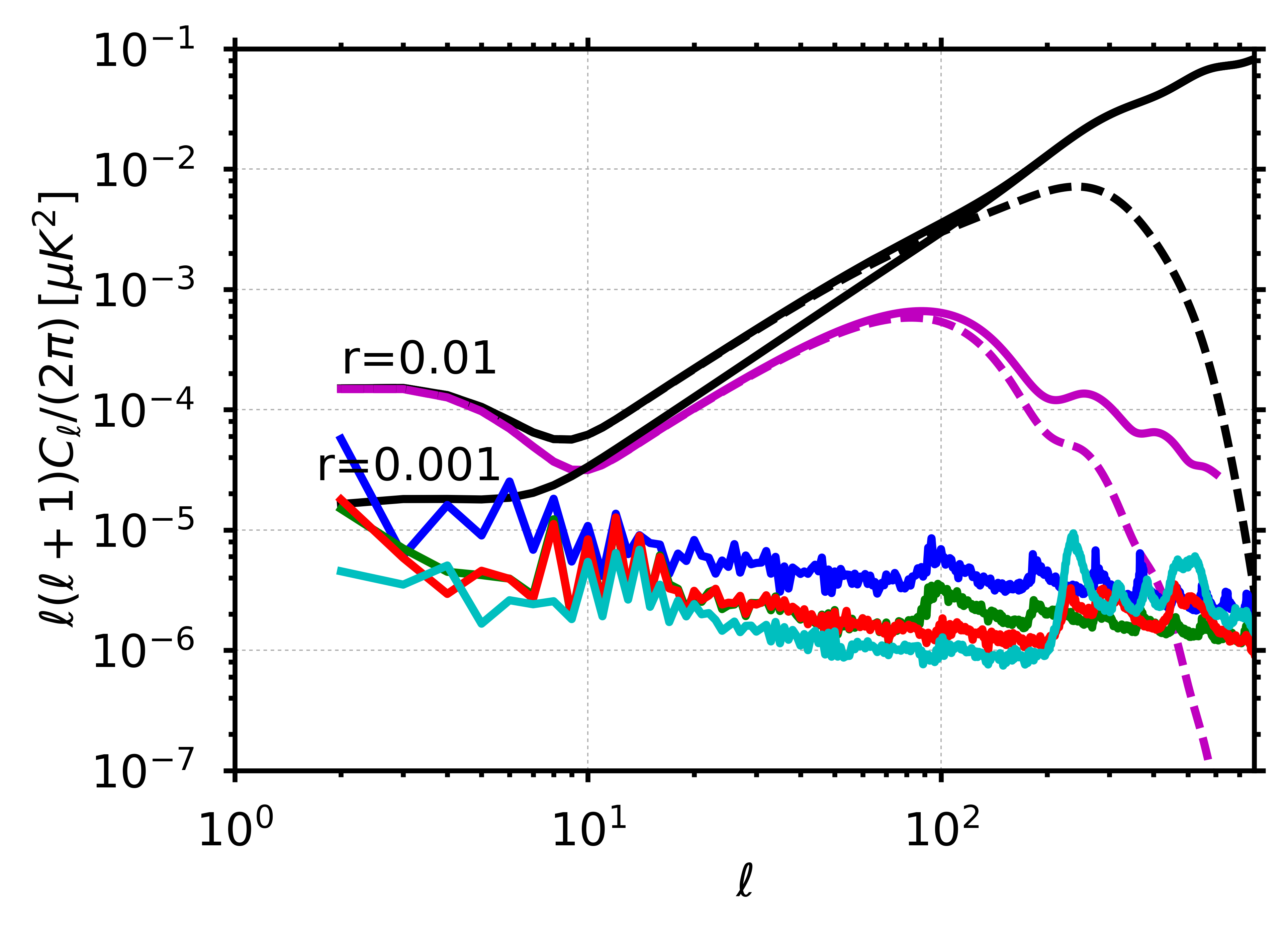}
\caption{$BB$ leakage power spectra
for different scanning parameters. In cyan: $\alpha=65^\circ$, 
$\beta=30^\circ$, $\tau_{\rm spin}$=10\,min, $\tau_{\rm prec}$=96.1803\,min,
red: $\alpha = 50^\circ, \beta = 45^\circ$, $\tau_{\rm spin}$=10\,min, $\tau_{\rm prec}$=96.1803\,min,
green: $\alpha = 50^\circ$, $\beta = 45^\circ$, $\tau_{\rm spin}$=2\,min, $\tau_{\rm prec}$=4 day,
blue: $\alpha = 30^\circ, \beta = 65^\circ$, $\tau_{\rm spin}$=2\,min, $\tau_{\rm prec}$=4 day.
Spectra are computed for 222 detectors. Curves for the $B$ mode model are described in 
Fig.~\ref{SpectrLeak1overN} caption. For the scanning strategies with a long 
precession period, we computed spectra for 100 detectors rescaling
to 222 equivalent detectors using the $1/N_{\rm det}$ dependance.}
    \label{SpectrLeak1All}
\end{figure}

Figures~\ref{SpectrLeak1overN}, \ref{SpectrLeak0.3-0.1} and \ref{SpectrLeak1All} show 
the bandpass mismatch leakage contributions to the $EE$ and $BB$ power spectra in 
different observing configurations. The power spectra are computed after
the 20\%  of the sky where the thermal dust emission is  strongest is masked. 
The data in this masked region is set to zero with no apodization (which is
unnecessary since the small-scale power in the leakage maps dominates
over the spurious power induced by the masking). For comparison we also show the CMB B-
mode spectrum for two different values of $r$. The dashed curves show how the signal
is attenuated by beam smearing assuming the 140\,GHz FWHM beamwidth
of 32' fitted to a Gaussian profile for the present LiteBIRD configuration~\cite{Ishino16}.
As will be demonstrated later, neglecting the discreteness of the scans, the overall 
amplitude of the leakage due to bandpass mismatch is
nearly Gaussian and of zero mean, and 
the variations of $\gamma_{dust}$ impact all multipoles of the leakage 
map power spectrum in a correlated way. For this reason, an accurate estimate of the 
average leakage power spectrum requires averaging many independent realizations 
even if many detectors are used for the simulations. At 
least on large angular scales, the fluctuations in the power spectrum due to different 
realizations is roughly an overall amplitude varying as the square of a Gaussian.

We find that with all other parameters equal, the bandpass mismatch error amplitude 
scales as $1/\sqrt{N_{\rm det}}$ where $N_{\rm det}$ is the number of detectors (and 
hence the power spectrum scales as $1/N_{\rm det}$). This scaling becomes more accurate 
when $N_{\rm det}$ becomes large, as shown by comparing the $EE$ and $BB$ leakage power 
spectra for $\tau_{\rm spin}$ = 10\,min, $\tau_{\rm prec}$= 96.1803\,min and
$N_{\rm det}$ of either 74 or 222. The pairs of spectra have the same shape but the 
ratio of power spectrum amplitudes is consistent with the  predicted ratio $222/74=3.$

Figure~\ref{SpectrLeak0.3-0.1} shows the $BB$ power spectra for $\alpha = 65^\circ$, $\beta = 30^\circ$
for several spin and precession period combinations.
We see that the characteristics of the leakage power spectrum (and in particular the location of the peaks
at $\ell \leq 100$), depend on the exact values of $\tau _{\rm spin}$ and $\tau _{\rm prec}$. A proper value of the ratio 
$\tau _{\rm prec}/\tau _{\rm spin}$ moves the peaks in the bandpass leakage spectrum to higher $\ell$, away from the location of the maximum of the primordial B-mode recombination bump.

Figure~\ref{SpectrLeak1All} compares the $BB$ power spectra for different opening 
angles $\alpha $ and $\beta ,$ and also different scan rates. With the constraint
$\alpha + \beta = 95^\circ$, scan strategies with larger precession angle produce less 
leakage because they allow for more homogeneous scan angle coverage per pixel, and 
hence lower $\left | \left< \cos 2\psi_j \right> \right |$ and
$\left | \left< \sin 2\psi_j \right> \right |$ per individual detector.

We observe that the power spectra above (without a HWP) are approximately proportional 
to $\ell^{-\eta}$ where $\eta \approx 2.5$. This angular power spectrum is less steep 
than that of dust emission itself. The shape of the resulting leakage spectrum can be 
expressed as a kind of convolution between the harmonic coefficients of the crossing 
moment maps and of the dust component map (see Ref.~\cite{martinPaper} for an 
analytical explanation of this power law). This spectral shape is problematic on very 
large scales, for example near the reionization bump, because the ratio of the bandpass 
mismatch to the white noise component of the detector noise (having an $\eta \approx 0$ 
spectrum) increases toward lower multipole number $\ell.$ We observe some dependance of 
the amplitude of the leakage spectra with respect to the scanning strategy parameters
$\alpha$ and $\beta$. Scanning strategies with more uniform angular coverage (provided 
by larger precession angles for the studied cases) have a lower leakage amplitude.

When the experiment observes with a rotating HWP, 
the equivalent of an optimized polarimeter configuration
is straightforwardly obtained when the HWP observes a given sky position $\hat p$ 
during an integer number of turns (and, thus, for an evenly spread set of angles between 0 and 2$\pi$).
In practice however, the pointing direction moves while the HWP rotates, and hence
data samples are not usually so evenly distributed. However, when the HWP rotates 
at 1.467\,Hz (88\,rpm) while the instrument beam scans the sky with a spin period of $\tau_{spin}=10$\,minutes
and with a $30^\circ$ angle, the beam is displaced by $0.204^\circ$ (about $12.3^{\prime}$) each time 
the HWP makes one turn. Neglecting this displacement, single detector timelines of $I$, $Q$, and $U$
with no bandpass mismatch leakage can be straightforwardly obtained from the data set,
and projected onto sky maps with optimal noise averaging, i.e., equivalent to the generalized 
least square solution of eqn.~(\ref{mapMakingEquation}). Of course, a real-life HWP is not perfectly achromatic
and hence is likely to introduce bandpass mismatch effects of its own. We postpone to future work the study of this type
of effect.

\begin{figure}[ht]
  \centering
  \includegraphics[width=1\columnwidth]{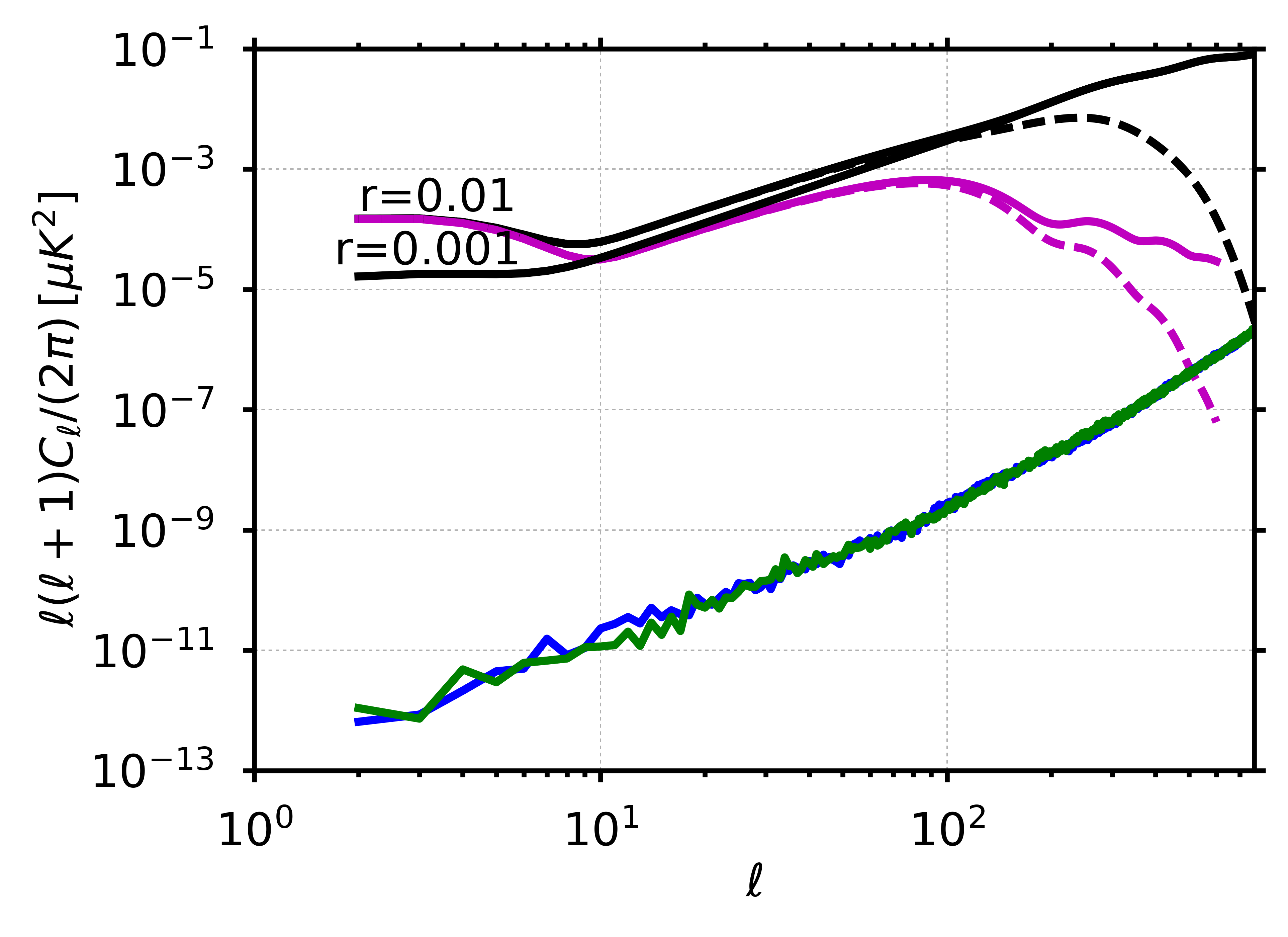}
  \caption{{\bf $EE$ and $BB$ leakage power spectra with rotating HWP}
    for $\alpha=65^\circ$, $\beta=30^\circ$ and spin period of
    10\,min with a HWP rotating at 88 rpm for
    50 detectors.}
    \label{SpectrLeakHWP}
\end{figure}

To illustrate the added value of a perfect HWP, we perform a simple set of 
simulations in which the input sky (smoothed by a $32^\prime$ beam) 
is a Healpix map pixelized at $n_{\rm side}=256$. The pixel size is well matched to 
the rotation speed of the HWP, which makes about one turn while it crosses a pixel. 
However, numerical effects will generate unevenness in the angular coverage of each 
pixel, and thus, when multi-detector maps are made using 
eqn.~(\ref{mapMakingEquation}), small bandpass leakage mismatch effects will subsist. 
Simulating the observation of this model sky with the use of a HWP spinning at 88\,rpm 
and other parameters set to $\alpha=65^\circ$, $\beta=30^\circ$, $\tau_{\rm spin}$
=10\,min, $\tau_{\rm prec}=96.1803$\,min, we obtain the small residual leakage shown in 
Fig.~\ref{SpectrLeakHWP}, which confirms the effectiveness of the HWP in reducing 
bandpass leakage by homogenizing the angular coverage in each pixel. The shape of the 
spectrum of the residual is similar to that of white noise. Its origin is in the small 
unevenness of the angle distributions across the pixels and is an artefact of sky 
pixelization.

We verify that in case of a perfect HWP, the multi--detector solution for the 
polarization is close to the solution consisting in combining single detector 
(including the HWP) polarization maps, as the residual leakage and its impact of $r$ 
that can be read off the plot, is negligible.

\begin{table}[b]
\centering
\begin{tabular}{|l|c|c|}
\hline
  &$2 \leq \ell \leq 10$ & $10 \leq \ell \leq 200$ \\
\hline
$\alpha = 30^{\circ}$; $\beta=65^{\circ}$; $\tau _{\rm prec}=4$\,days; $\omega _{\rm spin}=0.5$\,rpm & 1.83 $\times 10^{-3}$ & 9.32 $\times 10^{-5}$ \\
\hline
$\alpha = 50^{\circ}$; $\beta=45^{\circ}$; $\tau _{\rm prec}=4$\,days; $\omega _{\rm spin}=0.5$\,rpm & 6.49 $\times 10^{-4}$ & 4.66 $\times 10^{-5}$ \\
\hline
$\alpha = 50^{\circ}$; $\beta=45^{\circ}$; $\tau _{\rm prec}=96$\,min; $\omega _{\rm spin}=0.1$\,rpm & 6.32 $\times 10^{-4}$ & 3.08 $\times 10^{-5}$ \\
\hline
$\alpha = 65^{\circ}$; $\beta=30^{\circ}$; $\tau _{\rm prec}=93$\,min; $\omega _{\rm spin}=0.1$\,rpm & 3.29 $\times 10^{-4}$ & 7.61 $\times 10^{-5}$ \\
\hline
$\alpha = 65^{\circ}$; $\beta=30^{\circ}$; $\tau _{\rm prec}=96$\,min; $\omega _{\rm spin}=0.1$\,rpm & 3.27 $\times 10^{-4}$ & 2.11 $\times 10^{-5}$ \\
\hline
$\alpha = 65^{\circ}$; $\beta=30^{\circ}$; $\tau _{\rm prec}=96$\,min; $\omega _{\rm spin}=0.3$\,rpm & 3.03 $\times 10^{-4}$ & 1.77 $\times 10^{-5}$ \\
\hline

\end{tabular}
\caption{Contribution of bandpass mismatch error to 
the tensor-to-scalar ratio $r$ 
computed according to eqn.~(\ref{eq:chi_fit}). The level of the bandpass leakage relative to primordial B-mode signals is acceptable at the angular scale of the recombination bump, but problematic in the reionization bump at $\ell \leq 10$. Scanning strategies with larger $\alpha$ and smaller $\beta$ perform better, as they provide more uniform angular coverage in each pixel.
}
\label{r_fit}
\end{table}

Table \ref{r_fit} shows the contribution to $r$ that would 
result from uncorrected bandpass mismatch based on its 
power spectrum averaged over many realizations, calculated using
the projection
\begin{equation}
{\hat \delta r} = 
\dfrac{\sum_
{\ell =\ell _{\rm min}}^{\ell _{\rm max}} 
(2\ell+1) C_\ell\hat C_\ell }
{\sum_{\ell =\ell _{\rm min}}^{\ell _{\rm max}} 
(2\ell+1) C^2_\ell }.
\label{eq:chi_fit}
\end{equation} 
Here $C_\ell $ is the power spectrum for the primordial $B$ mode signal
normalized to $r=1.$ The Table shows $\delta r$ calculated for two ranges of $\ell $:
one with $\ell \in [2,10]$ to isolate the signal 
from the re-ionization bump, and another with $\ell \in [10,100]$ to
isolate the signal arising from the recombination bump. The
results in the table assume $N_{\rm det}=222$ detectors, but can be rescaled based 
on the $1/N_{\rm det}$ dependence to other numbers of detectors.
These results are only an order of magnitude estimate because they are
based on a single 140\,GHz channel, and it has been assumed that
very low and very high frequency channels have been used to 
removed the non-primordial components completely. We stress
that the bandpass mismatch power spectrum is not a simple bias
that can be predicted and subtracted away because its overall
amplitude suffers large fluctuations, which is of the same order 
of magnitude as the average bias itself.
%A more complete discussion of the implications for detecting $r$ 
%is given in the final `Concluding Remarks' Section.

\subsection{Analytic estimates}
\label{sub:analytic}

With the objective of finding fast and easy ways to predict the
magnitude of potential leakage without running many Monte Carlo
simulations, and in order to understand how the patterns shown in the
leakage map in Fig.~\ref{MapLeak} are related to the scanning
strategy, we now study theoretically in more detail how the leakage 
manifests itself in the polarization maps. To this effect, we expand the solution of
the map making equation [eqn.~(\ref{mapMakingEquation})].

We derive a simple expression for the leakage originating from differencing the 
signal from a pair of orthogonally polarized detectors observing instantaneously 
at the same location in the sky, so that data of the two detectors of the pair $i$ at time $t$ 
in pixel $p$ denoted as $S_{i;a}(t)$ and $S_{i;b}(t)$ are given by
\begin{eqnarray}
  S_{i;a}(t) &=& I_{i;p} + Q_p\cos 2\psi (t)  + U_p\sin 2\psi (t)  + M_{i;p},\cr 
  S_{i;b}(t) &=& I_{i;p} - Q_p\cos 2\psi (t)  - U_p\sin 2\psi (t)  - M_{i;p}. 
  \label{eqModSb}
\end{eqnarray}
Here we assume no noise and perfect calibration on the CMB (e.g., using the CMB 
dipole), and $\psi$ is the polarizer angle for detector $a$. $I_{i;p},$ $Q_p,$ $U_p$ 
are the Stokes parameters of the sky signal, $I_{i;p}$ being the mean intensity 
parameter for the detector pair $i$, and $M_{i;p}$ represents the bandpass mismatch component, which is given by 
\begin{equation} 
M_{i;p}=\frac{1}{2}\sum _{(c)} \left( \gamma _{(c)}^a- \gamma _{(c)}^b \right) ~I_{p,(c)}.
\label{EQDefMp}
\end{equation} 
Here the index $(c)$ labels the non-CMB sky components. The coefficient
differences $\big( \gamma _{(c)}^a -\gamma _{(c)}^b\bigr) $
vary from detector pair to detector pair, as explained in Sect.~2 (see in particular 
eqn.~(\ref{eq:sumcomp})). To minimize clutter, we have suppressed the index $i$
labelling the detector pairs. We neglect the subdominant effect of bandpass mismatch on the 
polarized sky components. As in the previous Section, we neglect noise in our analysis. 
The estimated noiseless Stokes parameter maps ${\widehat{Q_p}}$ and ${\widehat{U_p}}$ can be
expanded as ${\widehat{Q_p}} = Q_p + \delta Q_p$ and ${\widehat{U_p}} = U_p + \delta U_p$,
where $\delta Q$ and $\delta U$ represent the leakages to 
polarization resulting from bandpass mismatch. 
Ideal solutions with no leakage are given in eqn.~(\ref{eqModSb}). 

The map making equation gives
\begin{eqnarray}
\begin{pmatrix}
\phantom{\Big|}\widehat{I}_{p} \\ 
\phantom{\Big|}\widehat{Q}_{p} \\ 
\phantom{\Big|}\widehat{U}_{p}
\end{pmatrix}
= 
\begin{pmatrix*}[l]
\phantom{\Big|}  1 \phantom{singe} &  0  & 0 \\ 
\phantom{\Big|}  0 &  \frac{1}{2} \left( 1 + \langle \cos 4\psi \rangle \right) & \frac{1}{2} \langle \sin 4\psi \rangle \\
\phantom{\Big|}  0 & \frac{1}{2}\langle \sin 4\psi \rangle & \frac{1}{2} \left( 1 - \langle \cos 4\psi \rangle \right) 
 \end{pmatrix*}^{-1}
\begin{pmatrix}
\phantom{\Big|} \langle S \rangle \\ 
\phantom{\Big|} \langle \frac{1}{2}(S_a - S_b)\cos 2\psi \rangle  \\ 
\phantom{\Big|} \langle \frac{1}{2}(S_a - S_b)\sin 2\psi  \rangle
\end{pmatrix},
\end{eqnarray}
and the zeros in the 3x3 matrix result because the exact orthogonality of the two detectors of each pair
insures that $\langle \cos 2\psi \rangle $ and $\langle \sin 2\psi \rangle $ vanish exactly [compare 
with eqn.~(\ref{bpmErrorMatEqn})], so that the expression for 
$\widehat{I}_{p}$ decouples from the expressions for $\widehat{Q}_{p}$ and $\widehat{U}_{p}.$
Consequently, the leakages are given by 
\begin{eqnarray}
&&
\begin{pmatrix}
   \delta Q_p\\
   \delta U_p
\end{pmatrix}=
\begin{pmatrix*}[l]
\frac{1}{2}(1+\langle \cos 4\psi ) \rangle & 
\frac{1}{2} \langle \sin 4\psi \rangle\\
\frac{1}{2} \langle \sin 4\psi \rangle & 
\frac{1}{2}(1-\langle \cos 4\psi ) \rangle  
\end{pmatrix*}^{-1}
  \begin{pmatrix}
   \langle M_p\cos 2\psi \rangle\\
   \langle M_p\sin 2\psi \rangle
 \end{pmatrix}
\phantom{\Bigg|}
\cr 
\phantom{\Bigg|}
&&\quad = 
\frac{2}{
\left(1 - \langle \cos 4\psi \rangle^2 - \langle \sin 4\psi \rangle^2 \right)
}
 \begin{pmatrix*}[l]
1 + \langle \cos 4\psi \rangle  & -{\langle  \sin 4\psi \rangle} \\
   -{\langle  \sin 4\psi \rangle}  & 1- \langle \cos 4\psi \rangle 
 \end{pmatrix*}
  \begin{pmatrix}
   \langle M_p\cos 2\psi \rangle\\
   \langle M_p\sin 2\psi \rangle
 \end{pmatrix}.
\end{eqnarray}
Assuming that 
$\langle \cos 4\psi  \rangle^2 + \langle \sin 4\psi \rangle^2 \ll 1$
(which is not so bad an approximation except very near the poles),
we obtain 
\begin{eqnarray}
\begin{pmatrix}
   \delta Q_p\\
   \delta U_p
\end{pmatrix}\approx 
2
  \begin{pmatrix}
   \langle M_p \cos 2\psi \rangle\\
   \langle M_p \sin 2\psi \rangle
 \end{pmatrix}.
  \label{Eq:MM2det}
\end{eqnarray}

\begin{figure}
  \centering
  \includegraphics[width=0.49\columnwidth]{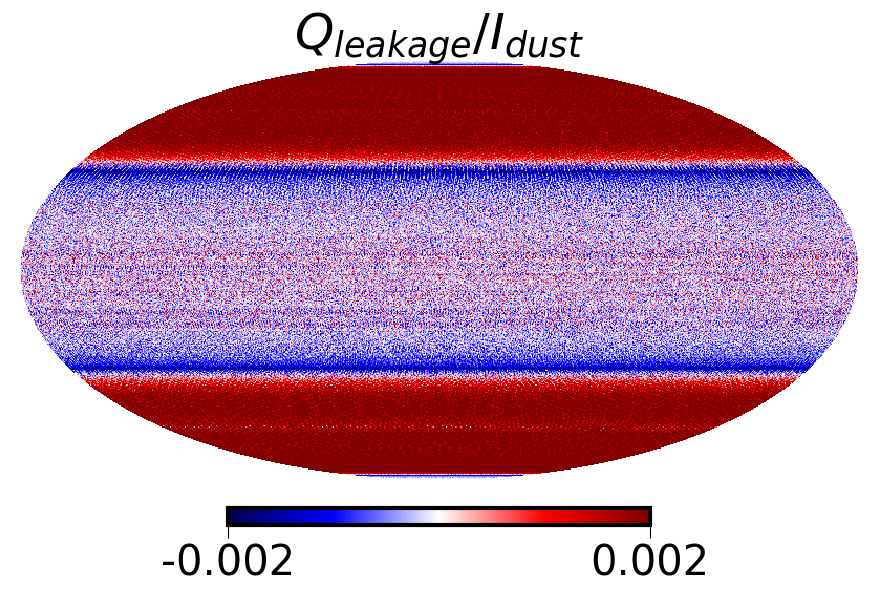}
  \includegraphics[width=0.49\columnwidth]{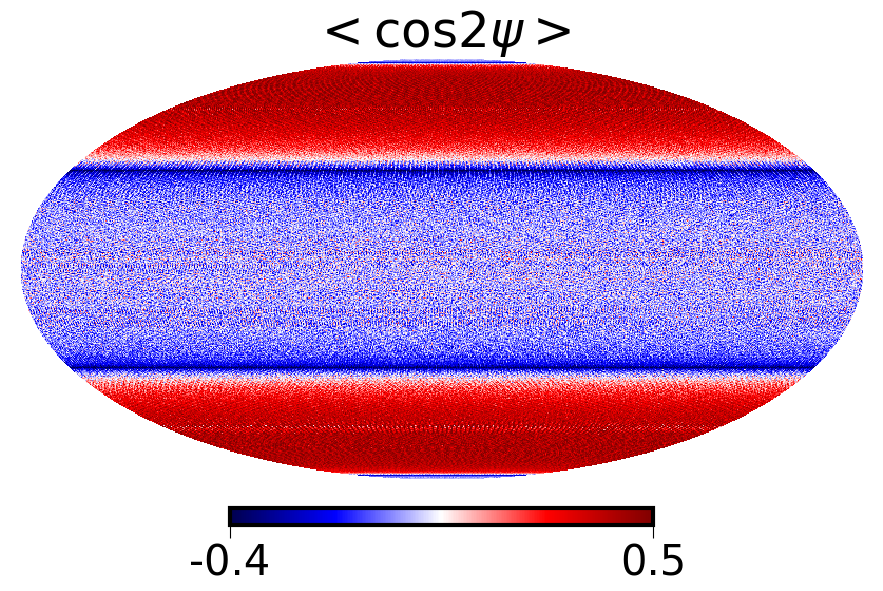}
  \caption{Left: Leakage for the $Q$ component relative to the dust
    temperature ($\delta Q/I_{{\rm Gal}}$) after polarization
    reconstruction using one bolometer pair only and a one year
    observation time. Right: Averaged $\cos 2\psi$ in each pixel for one bolometer
    after one year observation time. This quantity is strongly correlated 
    to the relative leakage $Q$ component with respect to the dust
    intensity.}
    \label{deltQoT}
\end{figure}

For one Galactic component, by replacing $M_p$ by its expression in eqn.~(\ref{EQDefMp}),
the relative amplitude of the leakage can be written as
\begin{equation}
\begin{pmatrix}
{\delta Q_p\over I_{{\rm Gal};p}}\\
{\delta U_p\over I_{{\rm Gal};p}}
\end{pmatrix}
=\left({\gamma _{\rm Gal}^a-\gamma _{\rm Gal}^b}\right) 
\begin{pmatrix}
\left< \cos 2\psi \right> \\
\left< \sin 2\psi \right> 
\end{pmatrix}.
\label{Eq:LeakToCrossLink}
\end{equation}
The term on the right hand side is one of the crossing moment terms for a single 
detector. We should then observe a large correlation between the two maps on the two 
sides of the equation. We have verified, with the help of simulations of data for one 
detector pair, this relationship for two different scanning strategies:
$\alpha=65^{\circ}$ and $\beta=30^{\circ}$ and $\alpha=50^{\circ}$ and
$\beta=45^{\circ}$. Figure~\ref{deltQoT} shows the relative leakage map
${\delta Q_p/I_{{\rm Gal};p}}$ and the quantity ${\sum \cos 2\psi/n_p}$. 
The $U$ component (not shown here) exhibits similar properties.

\begin{figure}
  \centering
  \includegraphics[width=1\columnwidth]{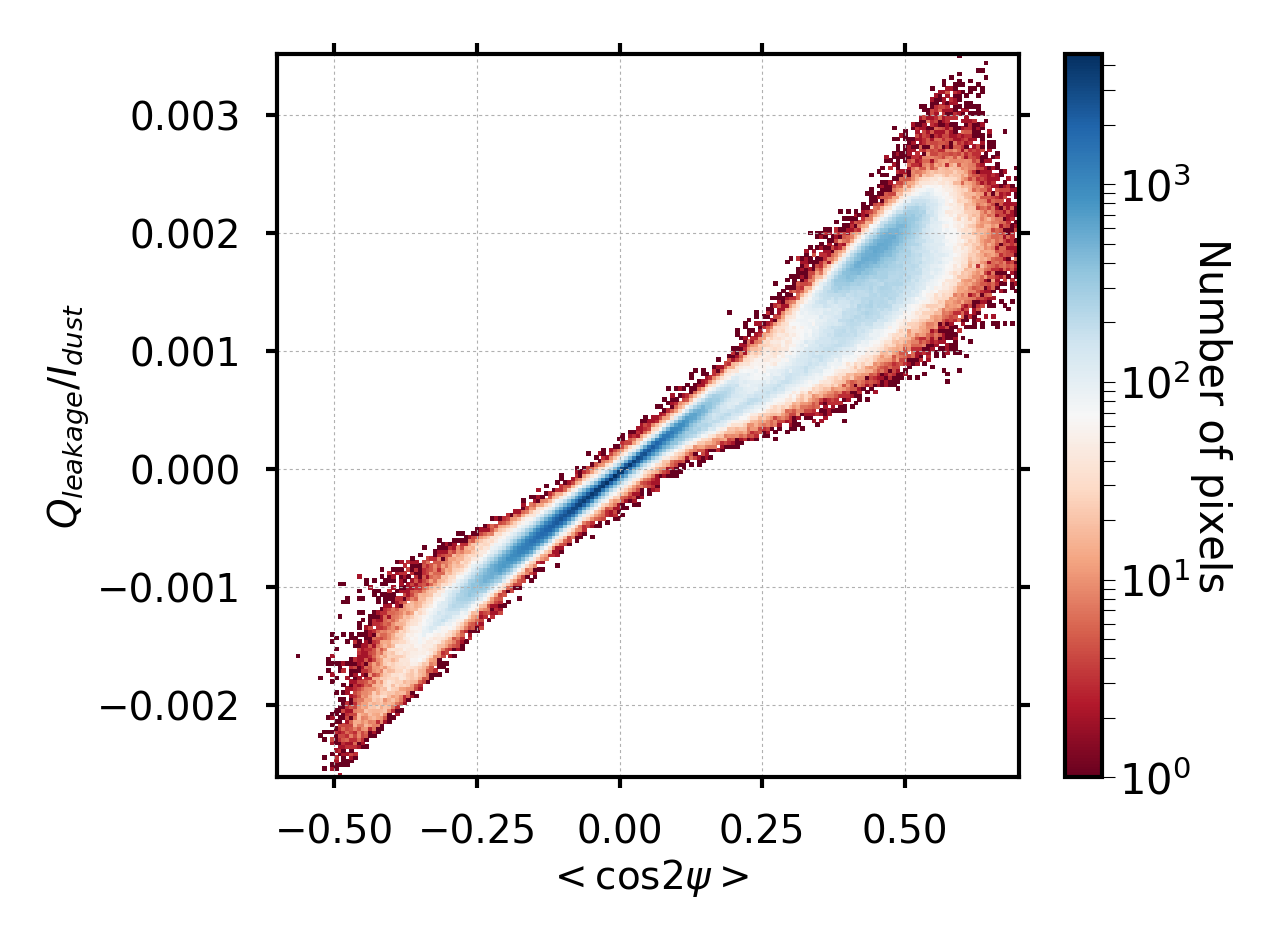}
  \caption{Values of the relative leakage ${\delta Q_p/I_{{\rm Gal};p}}$ for a pair of 
  detectors with orthogonal polarizations of a function of the scanning strategy 
  parameter $ (1/n_p) \sum \cos 2\psi $ (see text) after map making with two 
  detectors only. We observe a tight correlation between the relative leakage and the 
  second order crossing moments.}
    \label{scatter}
\end{figure}

Figure~\ref{scatter} shows the correlation of the two maps, by plotting the values of 
one map versus the other for a subset of pixels. We observe a high correlation between 
the two maps. We verify that the slope is given by the coefficient
$\Delta \gamma = \gamma_a-\gamma_b$ as derived in eqn.~(\ref{Eq:LeakToCrossLink}). This figure shows the 
tight link between the crossing moments and the relative leakage due to bandpass 
mismatch. It also shows that the approximations made to derive 
eqn.~(\ref{Eq:LeakToCrossLink}) are valid since we observe a relatively small scatter 
around the linear slope. The outliers in the Figure are due to pixels near the ecliptic 
poles where the angle coverage is less uniform for the scanning parameters used as a 
baseline in this work.

We now consider the solution combining more detectors. The generalization of eqn.~(\ref{Eq:MM2det})
gives for the resulting leakage component
\begin{eqnarray}
 \begin{pmatrix}
   \delta Q_p\\
   \delta U_p
 \end{pmatrix}=
 \begin{pmatrix}
   \frac{1}{2}\sum\limits_i \sum\limits_j (1 + \cos 4\psi_i^j) &    \frac{1}{2}\sum\limits_i\sum\limits_j \sin 4\psi_i^j\\
  \frac{1}{2}\sum\limits_i\sum\limits_j \sin 4\psi_i^j & \frac{1}{2} \sum\limits_i \sum\limits_j (1 - \cos 4\psi_i^j)
 \end{pmatrix}^{-1}
  \begin{pmatrix}
   \sum\limits_i\sum\limits_j \cos 2\psi_i^j \,M_{i,p}\\
   \sum\limits_i\sum\limits_j \sin 2\psi_i^j \,M_{i,p}
 \end{pmatrix}
 \label{Eq:MMmultidet}
\end{eqnarray}
where we sum over all the detector pairs indexed by $i$ and over all samples $j$ 
falling in pixel $p$ for each detector. In this case, for which we consider the 
realistic configuration of more than one pair of detectors per pixel, the covariance 
matrix above becomes nearly diagonal. As the number of detectors is increased, the 
matrix in eqn.~(\ref{Eq:MMmultidet}) becomes increasingly diagonal. The total leakage 
is then simply, replacing the leakage term $M_p$ by its expression:
\begin{equation}
  {\delta Q_p\over I_{{\rm Gal};p}}
\approx {2\over N_{\rm hit}} \sum_i \Delta \gamma_i \, \sum_j \cos 2\psi_i^j ,
\end{equation}
using eqn.~(\ref{EQDefMp}), where we have defined $N_{\rm hit}$ as the total number of 
hits including all detectors (and not only count 1 per detector pair which explains the 
cancellation of the 1/2 factors since the sum runs over detector pairs), and $\Delta\gamma_i = \gamma_i^a - \gamma_i^b$. The leakage 
vanishes if each individual detector has uniform angle coverage. We observe that 
the relevant quantities to estimate the level of leakage for a given scanning strategy 
are the individual detector second order crossing moments. Following our hypothesis 
that the $\gamma$ parameters are random and uncorrelated, we express the variance 
of the leakage map as:
\begin{equation}
 {\rm Var}\left( \delta Q_p\over I_{{\rm Gal};p}\right) 
\approx \sum_i {\rm Var}(\Delta \gamma_i)~\left(\sum_j \cos 2\psi_i^j
\right)^2\left({2\over N_{\rm hit}}\right)^2,
\end{equation}
which gives, since ${\rm Var}(\Delta \gamma ) = 2 {\rm Var}(\gamma )$:
\begin{equation}
 {\rm Var}\left( \delta Q_p\over I_{{\rm Gal};p}\right) 
\approx 4{{\rm Var}(\gamma )\over N_{\rm det}}\, 
\Bigg \langle{{\left({\sum \cos 2\psi_i^j\over {\bar n_p}}\right)^2}}\Bigg \rangle_{\rm det},
\label{varleak}
\end{equation}
where $\langle ~\cdot ~ \rangle_{\rm det}$ denotes average over all detectors, and
${\bar n_p} = {N_{\rm hit}\over {N_{\rm det}}}$ is the average number of hits per 
detector. The expression for the $U$ component is similar with the cosine replaced by a 
sine. This expression of the variance of the leakage map is also valid if detectors are 
not arranged by pairs.

\begin{figure}
  \centering
  \includegraphics[width=0.49\columnwidth]{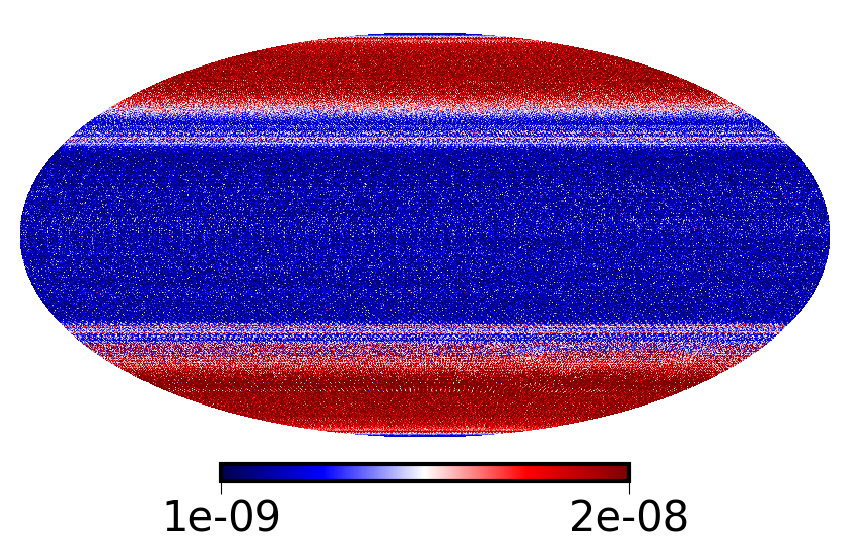}
  \includegraphics[width=0.49\columnwidth]{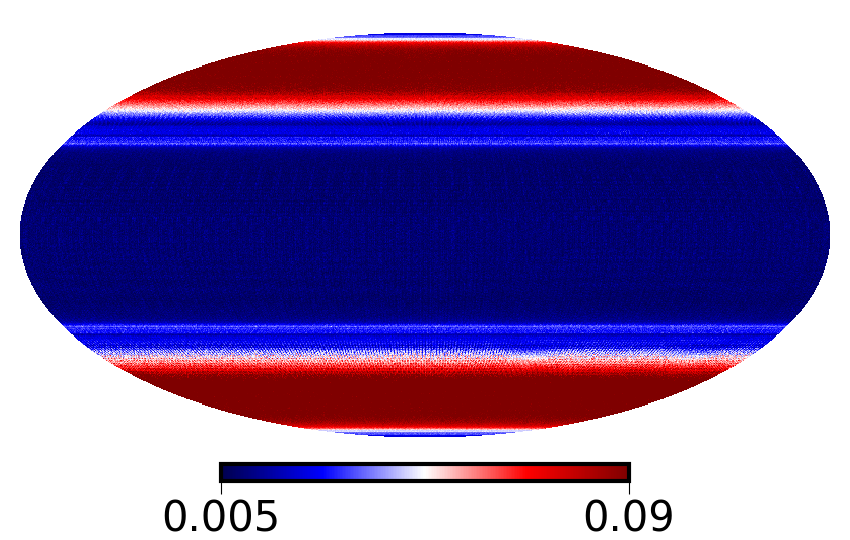}
  \caption{Left: Estimated leakage variance of the $Q$ component relative to 
  the dust temperature (${\rm Var}\left( \delta Q_p/I_{{\rm Gal};p}\right)$) after polarization reconstruction using
    all bolometer pairs and one year of observations. We used 10 independent 
    realizations of the bandpass to estimate the variance.
    Right: Averaged $\Big \langle \Bigl( 
(1/ {\bar n_p}) \sum \cos 2\psi_i^j \Bigr) ^2\Big \rangle _{\rm det}$ in each pixel
    for all bolometers after one year observation time. As for the detector pair case, 
    we observe a tight correlation of the two maps on large angular scales 
}
    \label{var_deltQoT}
\end{figure}

\begin{figure}
  \centering
  \includegraphics[width=1\columnwidth]{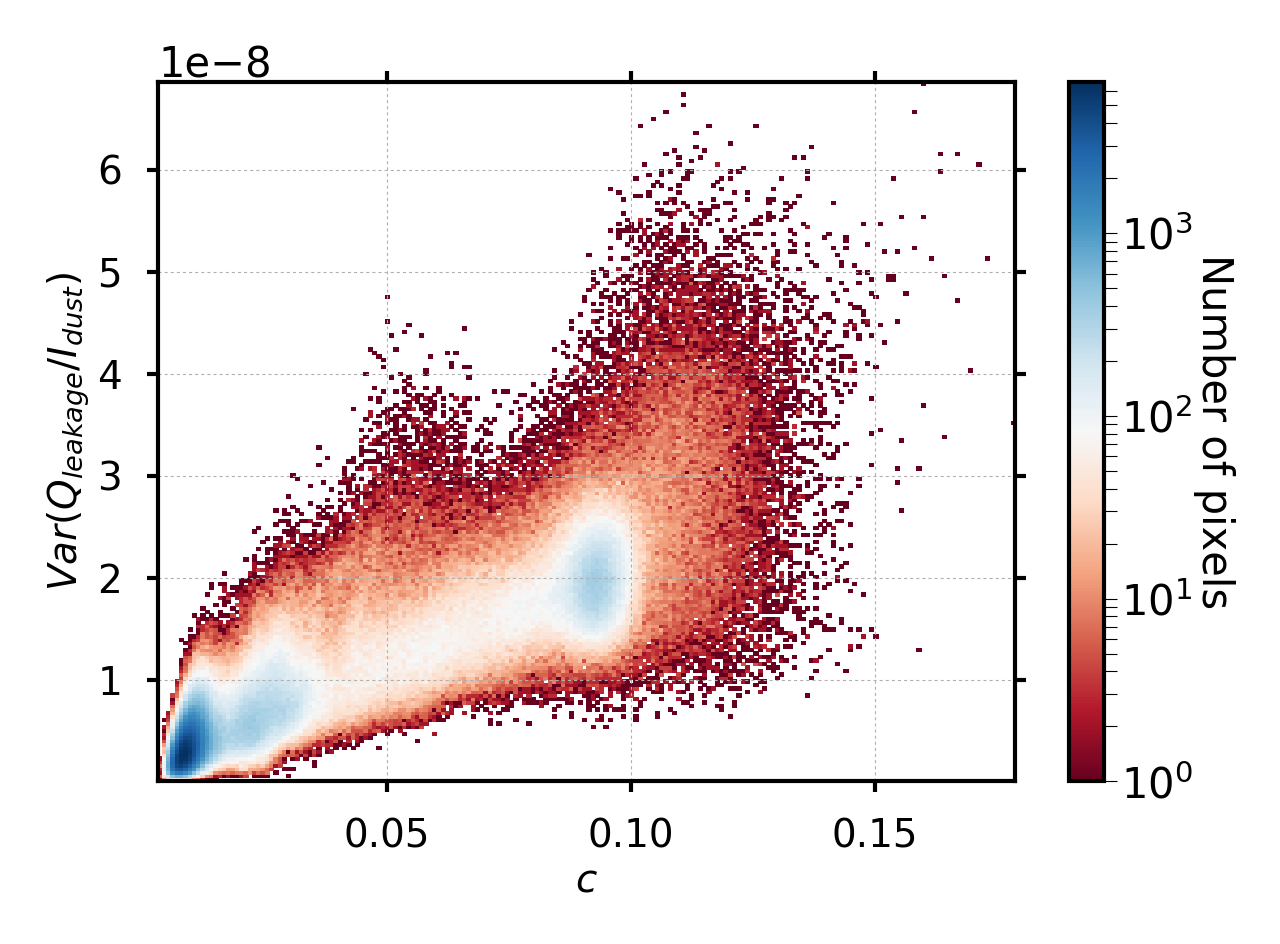}
  \caption{Estimated variance distribution 
     of the relative leakage parameter:
  ${\rm Var}\left( \delta Q_p/I_{{\rm Gal};p}\right)$ 
as a function of 
    $ c = \Big \langle
\Big(
(1/{\bar n_p})
\sum \cos 2\psi_{t;i}
\Bigr)^2
\Big \rangle_{\rm det}$ (see text) after map making including
    all detectors. We have averaged over ten realizations to estimate the variance.}
    \label{scatter_all_detectors}
\end{figure}

Figure~\ref{var_deltQoT} compares the maps of the variance on the left-hand side of the 
previous relationship which was estimated with ten independent realizations of the 
bandpass parameters, with the quantity 
$\Big \langle{{\left(\phantom{\Big|}(1/{\bar n_{p}})\sum \cos 2\psi_i\right)^2}}\Big \rangle_{\rm det}$. Figure~\ref{scatter_all_detectors} 
shows the correlations between the two quantities on a scatter plot. We observe a 
significant correlation of the two quantities, especially on large scales. The 
dispersion is partly due to the limited number of realizations to estimate the 
variance. Nevertheless, this shows that the level of leakage can be evaluated by 
estimating the second order crossing moments only for different scanning strategies 
without the need of running large simulations. This result explains what was observed 
in Fig.~\ref{SpectrLeak1All}, showing the level of the leakage with respect to the 
scanning parameters $\alpha$ and $\beta$. The strategies with more uniform angle 
distribution (the ones with larger precession angle) show lower residuals (see
also~\cite{Wallis16} for the link with other systematic effects).

% BOO

Results show that contamination from bandpass mismatch even if small could
contribute to the $B$ mode spectrum at a non-negligible level, close to the detection 
limit of primordial $B$ modes with future satellite missions. Systematic
variation of the bandpass functions across the focal plane, as opposed to
the uncorrelated random variations assumed in this study, could produce
larger errors. These considerations motivate developing correction methods, which we 
present in the companion paper~\cite{BPMMII}.

\subsection{Importance of avoiding resonances}
\label{sub:resonances}

Here we briefly explain some considerations for choosing the scan frequency parameters 
$\omega _{\rm spin}$ and $\omega _{\rm prec}.$ We found that to obtain good
crossing moment maps, careful attention must be paid to choosing the ratios of the 
hierarchy of scan frequencies $\omega _{\rm ann}\ll \omega _{\rm prec}\ll \omega _{\rm spin},$
and when there is a continuously rotating HWP also $\omega _{\rm HWP}.$
For $\omega _{\rm prec}/\omega _{\rm ann},$ we choose to make this number an integer so 
that the scan pattern closes. In all the simulations reported here, we assumed a single 
survey of exactly one year in duration. Given the large number of precession cycles in
a year, this requirement can be achieved by means of a very small
adjustment in $\omega _{\rm prec}.$ One might also want to do the
same for the spin period, but this is less critical because
of its shortness compared to a year.

More critical is the ratio $\theta =\omega _{\rm spin}/\omega _{\rm prec},$
which must be chosen so that $\theta $ cannot be well approximated
by simple fractions of the form $p/q$ where $p$ and $q$
are relatively prime and $q$ is small in a sense that we shall make more precise
shortly. Of concern are exact or near exact resonances where
$q$ is less than of order $\omega _{\rm prec}/\omega _{\rm spin}.$

\begin{figure}[t]
\begin{center}
\includegraphics[width=0.49\textwidth ]{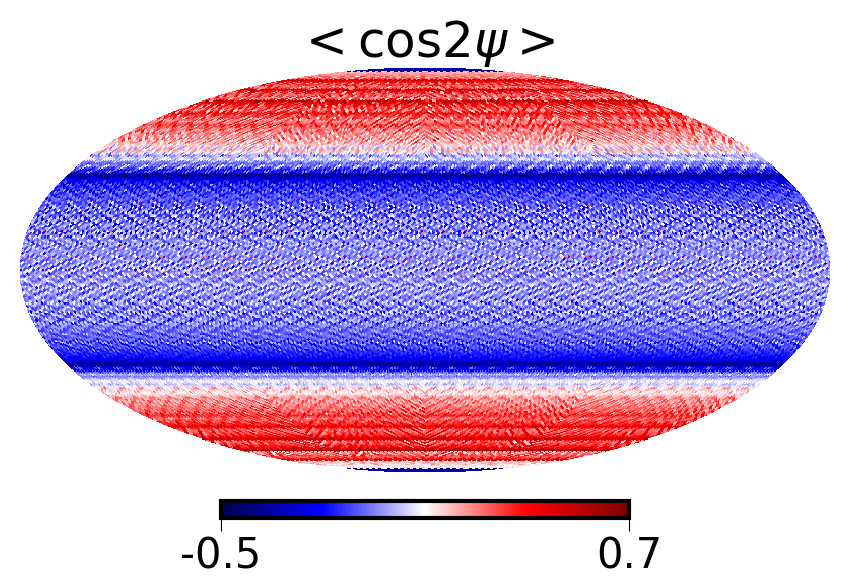}
\includegraphics[width=0.49\textwidth ]{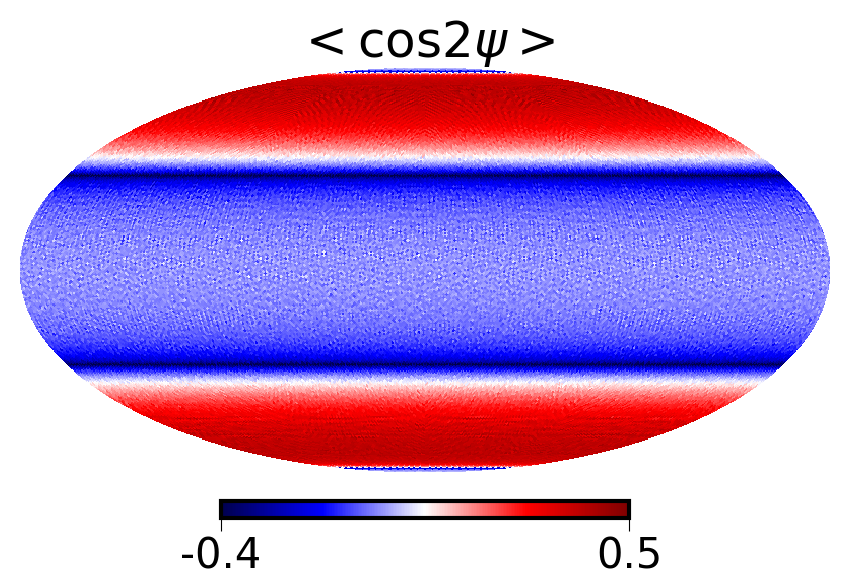}  
\end{center}
\caption{
{\bf Effect of a poorly chosen scanning frequency ratios.}
The map on the left has $\theta =\omega _{\rm spin}/\omega _{\rm prec}=9.3,$
whose continued fraction representation is $[9,3,3],$ whereas the lower map has the 
more irrational ratio $\theta =9.61803,$ whose continued fraction representation is
$[9,	1,1,1,......].$ A series of Moir\'e patterns on intermediate angular scales
is clearly visible in the map on the left, which lead to spikes in the crossing moment 
map power spectra, and also in the final bandpass mismatch power spectra. 
The artefacts can be avoided by choosing ratios of frequencies 
judiciously in order to avoid good rational approximations.}
\label{Fig:MoirePatterns}
\end{figure}

Before entering into the theory of how the ratio $\theta $ should
be chosen (and jumping ahead slightly), we show what goes wrong
when $\theta $ is not well chosen. For example, our first
try had $\tau _{\rm spin}=10$ min and $\tau _{\rm prec}=93$ min and gave
hit count and crossing moment maps with clearly visible Moir\'e patterns
at intermediate angular scales, as shown in Fig.~\ref{Fig:MoirePatterns}, 
showing clear evidence of a near resonance. However, when $\omega _{\rm prec}$
was sped up by the Golden ratio $\Phi =(1+\sqrt{5})/2=1.61803398875$
(reputed to be the ``most irrational'' number),\footnote{See for example
Michael Berry, (1978, September), {\it Regular and irregular motion}, in S. Jorna 
(Ed.), AIP Conference proceedings (Vol. 46, No. 1, 16-120), AIP for a nice discussion
of these questions in a different context, that of perturbations of integrable
systems in classical mechanics, KAM theory, and the stability of the solar system.}
these undesirable Moir\'e patterns disappear, as shown in the bottom right panel of the 
Figure. The same effect could be achieved by altering the ratio $\theta $
by just 5\%, so that the spin cycle has the same phase as with the Golden ratio sped 
up. We note that the effect of these Moir\'e patterns on the bandpass mismatch power 
spectra is to introduce peaks at multipole numbers at which the 
bandpass mismatch error is increased by up to about an order of magnitude beyond the 
baseline, where it would be if $\theta $ had been well chosen to avoid near resonances.
We also note that when a continuously rotating HWP is introduced, there are two 
independent ratios to worry about, although the artefacts are less acute than in the 
case of no rotating HWP.

The theory of choosing ratios to avoid near resonances relates to problems well studied 
by pure mathematicians in the area of number theory, or more specifically 
the theory of Diophantine approximations, and we discussed these issues in more detail 
elsewhere \cite{MartinNumTheory}. The tool for characterizing
the near resonance properties of real numbers is the continued
fraction representation, where we expand
\begin{equation}
\theta = [a_0, a_1, a_2, \ldots ]= a_o+\dfrac{1}{a_1+\dfrac{1}{a_2+\ldots }}
\end{equation}
where $a_0$ is an integer and $a_1, a_2, \ldots $ are positive
integers. For a rational number, the continued fraction representation
terminates; for an irrational number it is of infinite length.
The partial sums, known as `convergents,' generate a sequence of
`best rational approximations' $p/q$ to $\theta ,$\footnote{An 
irreducible fraction $p/q$ is a `best approximation' to $\theta $ if
$\vert \theta -p'/q'\vert >\vert \theta -p/q\vert $ whenever $q'<q.$}
with $q$ ascending. When a coefficient $a_n$ is large compared to one, the
preceding convergent is a particularly good approximation to $\theta $
considering the magnitude of $q.$ The Golden ratio $\Phi $ has the
continued fraction representation $[1,1,1,\ldots ],$ and thus
has among the worst approximation properties of any number.

\begin{figure}[h]
\begin{center}
\includegraphics[width=0.50\textwidth ]{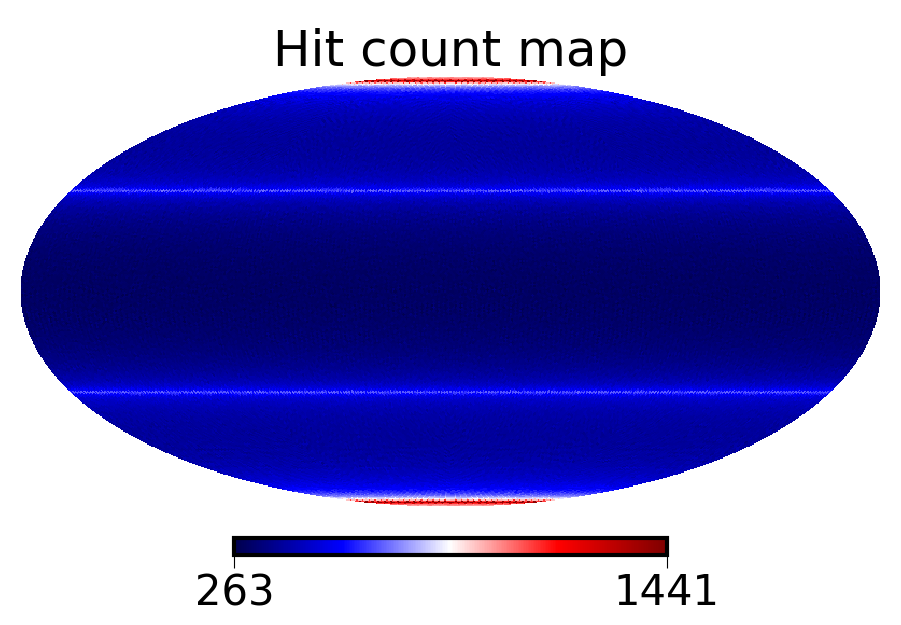}
\includegraphics[width=0.49\textwidth ]{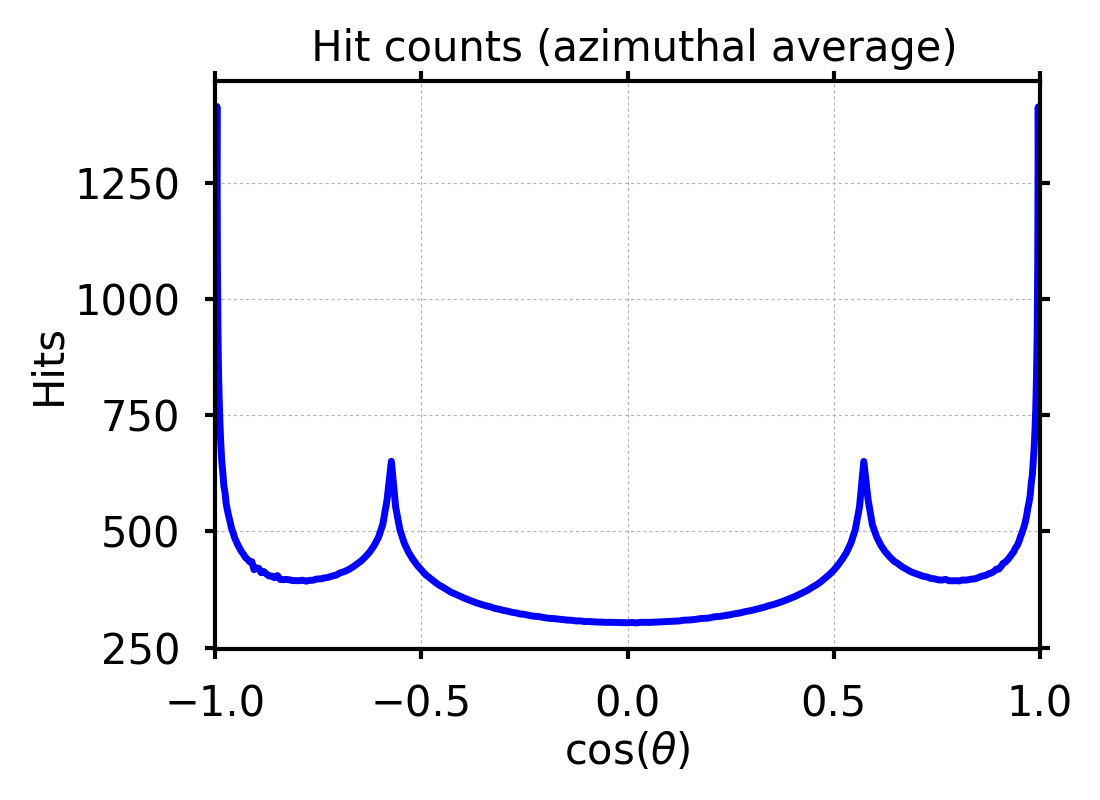}
\end{center}
\caption{{\bf Hitcount map and azimuthal average for fiducial scanning pattern}
The hitcount map is roughly uniform except for some localized spikes of high
density around the ecliptic poles and at the caustics at ecliptic latitude
$\pm(\alpha -\beta )=\pm (65^\circ -30^\circ )=\pm 35^\circ .$ In the bottom
plot the horizontal axis is $\cos \theta $ where $\theta $ is the angle from
the north ecliptic pole.
}
\label{Fig:HitcountMaps}
\end{figure}

For the parameters used in the simulations reported below, we 
adjusted the precession period so that there are an integer number 
5467 cycles in a sidereal year, giving a precession period of 
96.2080  minutes, and we replaced the ratio of $\theta =9.3,$ 
which in terms of continued fractions is represented by $[9,3,3],$ 
with the ratio 9.618033988749895,\footnote{In any specific application, the objective of avoiding near resonances
obviously requires an accuracy involving only a finite number of terms of the
continued fraction expansion. Moreover, it is less the instantaneous ratio
of frequencies that matters but rather the relative phase. We have found
using numerical simulations that avoiding Moir\'e patterns is achieved when
the ratios are maintained with a relative accuracy of 1 part in $10^3,$ although
the exact accuracy needed will depend on the particular application.} 
whose continued fraction representation is 
$[9,1,1,1,......],$ giving a spin period of 10.002876 minutes. 
One may ask: approximately to what accuracy would one wish to 
maintain this ratio? Certainly more accuracy than the inverse 
of the number of precession cycles in a year would be superfluous. In fact, 
less accuracy would be adequate, the  exact number depending 
on the precise scanning parameters, but we postpone further assessment of the required precision to future work. Moreover, it is more the absolute pointing that 
matters and not so much a question of maintaining precise ratios at any particular moment.

An important practical question is what accuracy is required in the ratios of the frequencies in 
order to avoid the Moir\'e patterns due to near resonances. It is not possible to provide 
a general answer to this question, but we performed some numerical experiments for the scanning 
frequencies considered in this paper and found that tuning the ratio of the frequencies to 
about 0.2\% sufficed. It should be stressed that it is the relative phase rather than the instantaneous ratio of frequencies that matters for avoiding Moir{\'e} artifacts. In the above discussion we considered only a single ratio, but for more complicated
situations with several frequencies, there is more than one ratio to keep away from 
near resonant values. A rotating half-wave plate, for example, introduces another frequency, and
in principle the annual drift also allows other dimensionless ratios of frequencies to be formed. 
These complications will be investigated elsewhere.

\begin{figure}[h]
\begin{center}
\includegraphics[width=0.49\textwidth ]{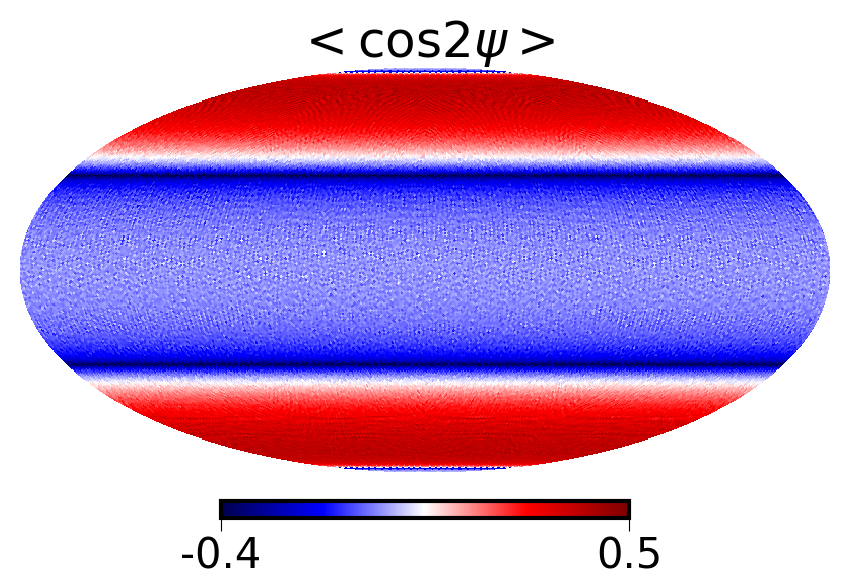}
\includegraphics[width=0.49\textwidth ]{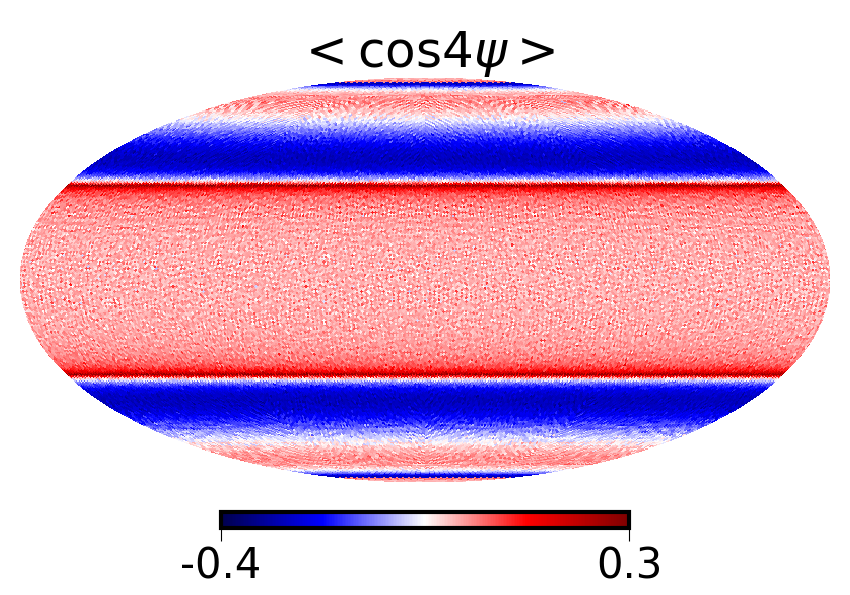}
\includegraphics[width=0.49\textwidth ]{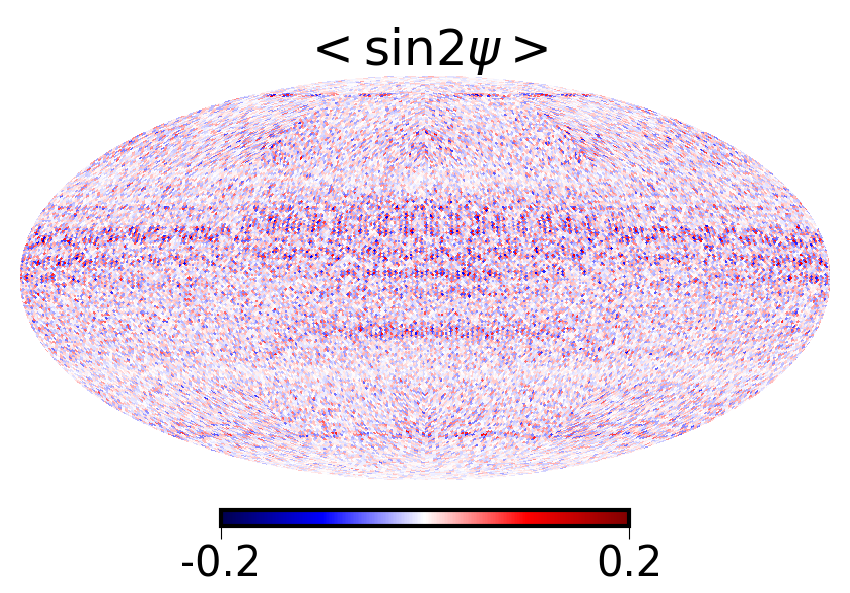}
\includegraphics[width=0.49\textwidth ]{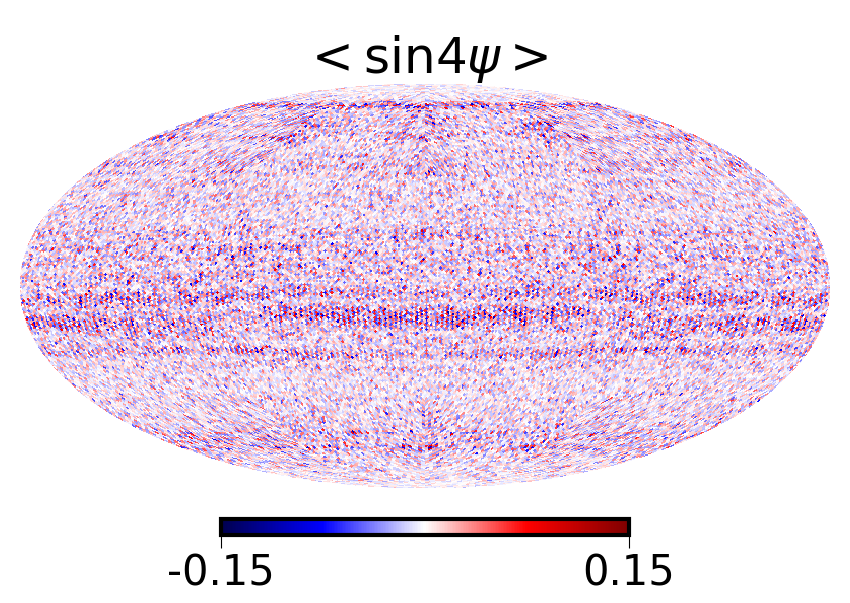}
\end{center}
\caption{{\bf Crossing moment maps for the fiducial scanning pattern.}
The four relevant crossing moment maps 
$\left< \cos 2\psi \right> ,$
$\left< \cos 4\psi \right> ,$
$\left< \sin 2\psi \right> ,$
$\left< \sin 4\psi \right> $
(left to right, top to bottom)
are shown for the fiducial scanning pattern (defined in the text) for a single detector 
whose polarization axis is oriented along the line running from the center of the beam 
to the spin axis. The corresponding maps for other polarizer orientations can be 
obtained trivially using the property that the first two maps transform as a spin-2 
vector and the second two as a spin-4 vector under rotations of the polarization 
orientation. We observe that the cosine maps have structures coherent on large scales 
and azimuthally symmetric in ecliptic coordinates, whereas the sine maps include only 
small-scale noise (which is also present in the cosine maps) but have no structure 
coherent on large angular scales.
}
\label{Fig:CrossLinkingMaps}
\end{figure}

\begin{figure}[h]
\begin{center}
\includegraphics[width=0.49\textwidth ]{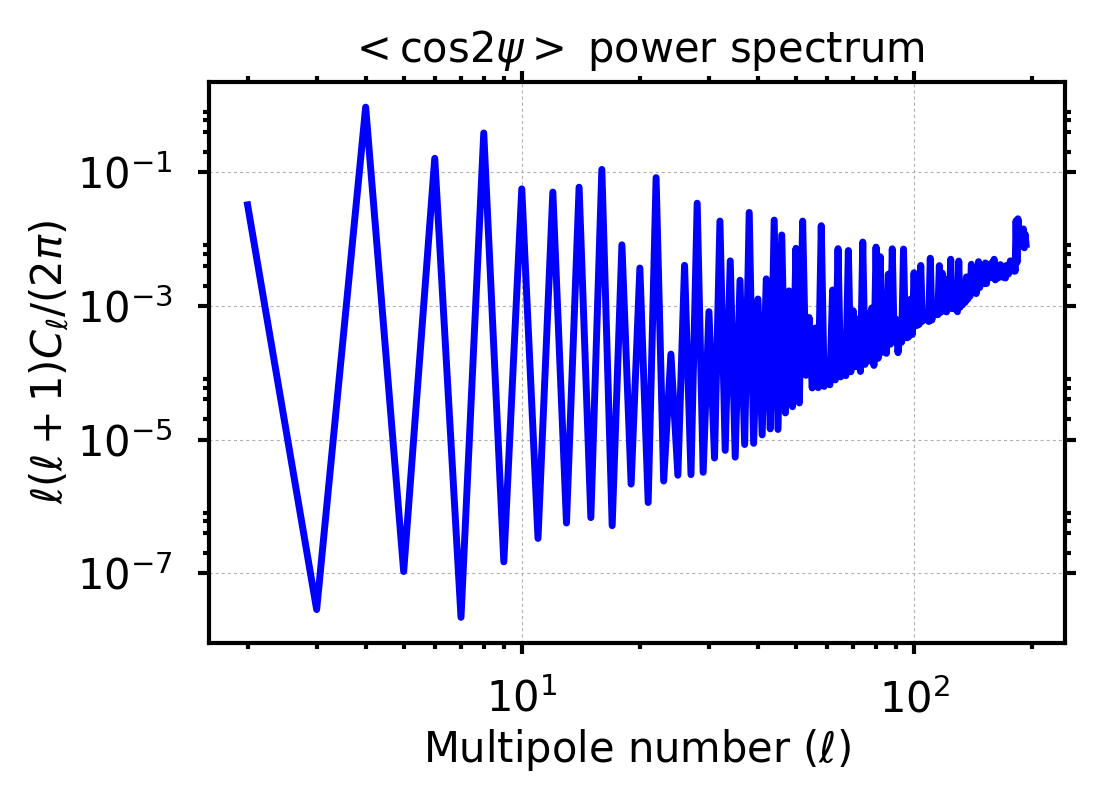}
\includegraphics[width=0.49\textwidth ]{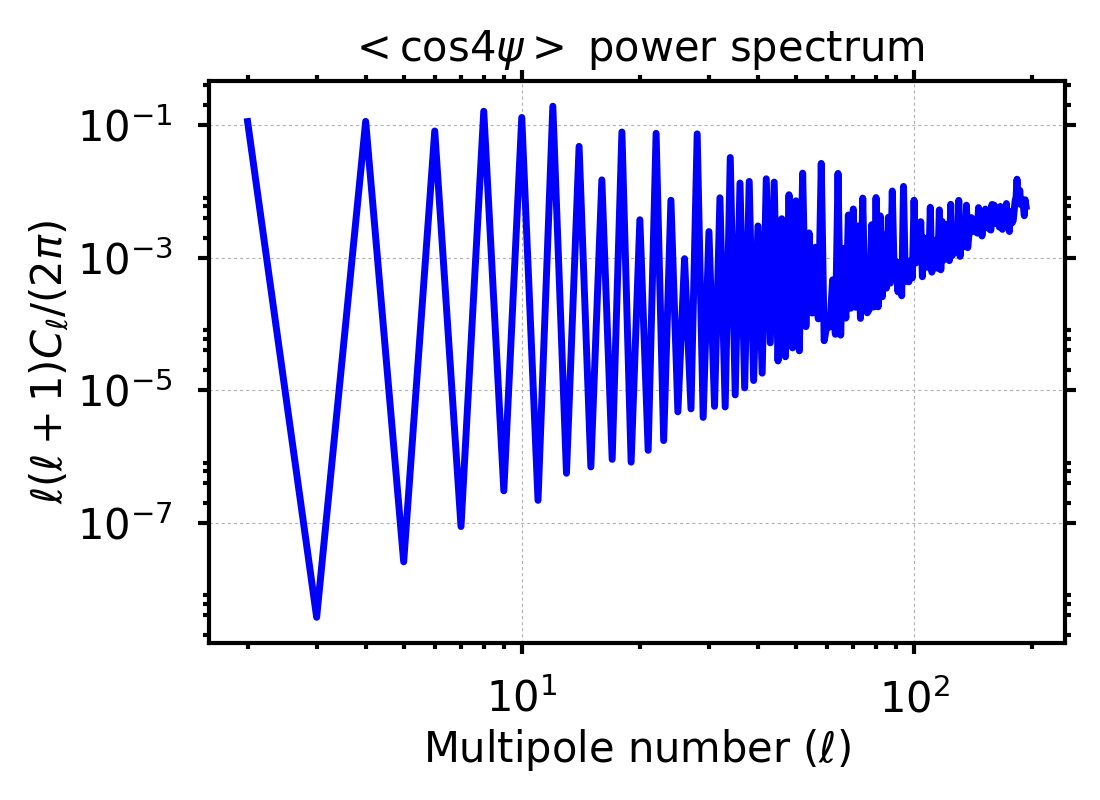}\\
\includegraphics[width=0.49\textwidth ]{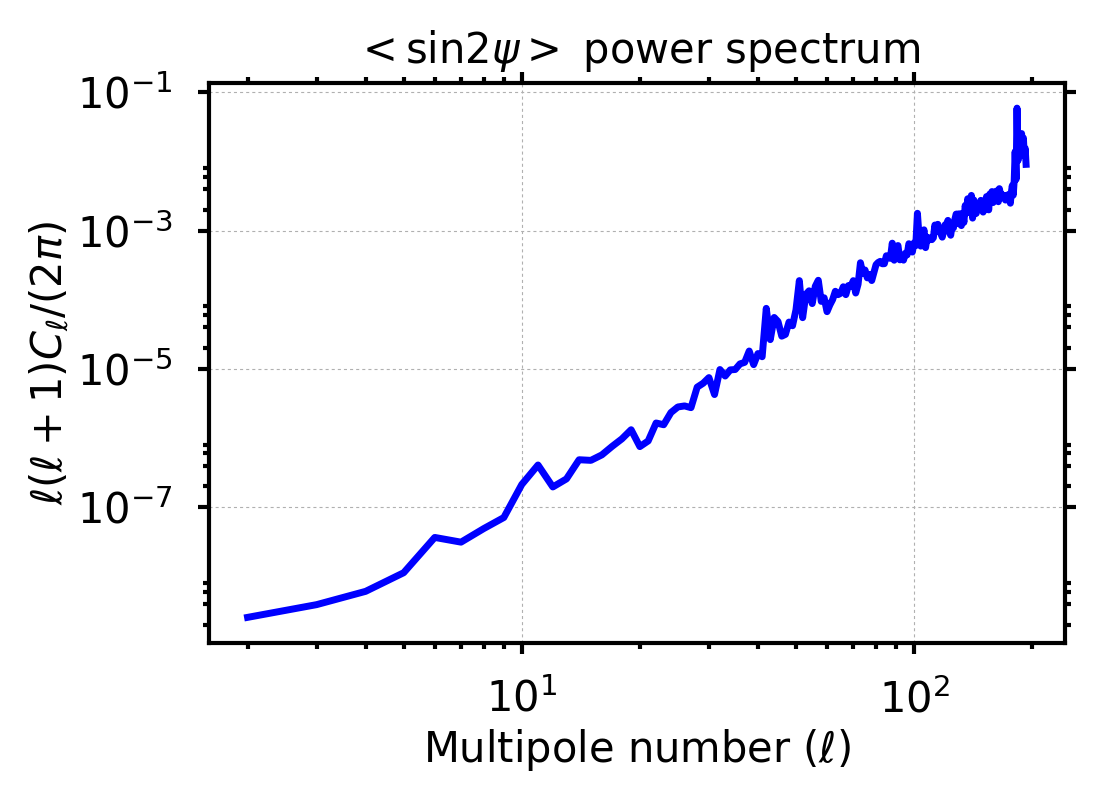}
\includegraphics[width=0.49\textwidth ]{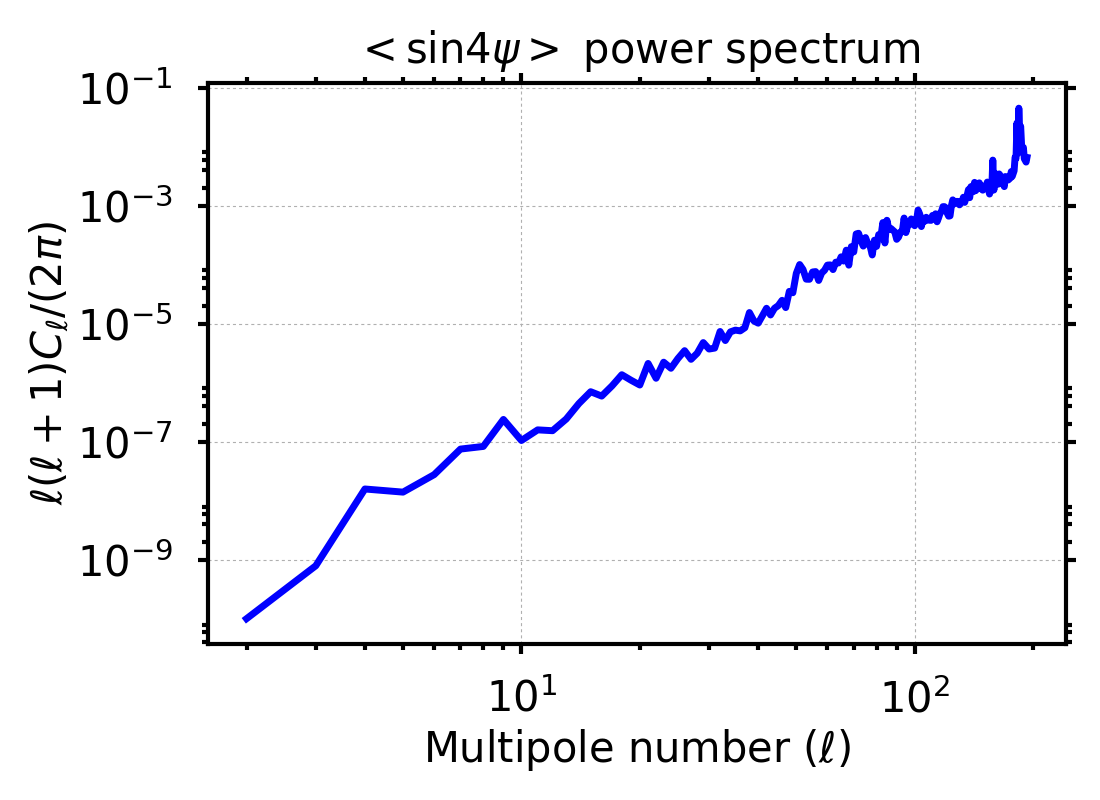}
\end{center}
\caption{{\bf Crossing moment map power spectra.} 
We show the power spectra of the maps of Fig.~\ref{Fig:CrossLinkingMaps}.
The spectra of the two cosine maps, because
of the azimuthally symmetric large power coherent on large scales shown in
Fig.~\ref{Fig:CrossLinkingAzimuthalAverages}, have 
power spectra scaling similar to $\ell ^{-2}$ for the even moments,
whereas the two sine maps (bottom) row exhibit spectra resembling pure white noise. 
}
\label{Fig:CrossLinkingPS}
\end{figure}

\begin{figure}[ht]
\begin{center}
\includegraphics[width=0.35\textwidth ]{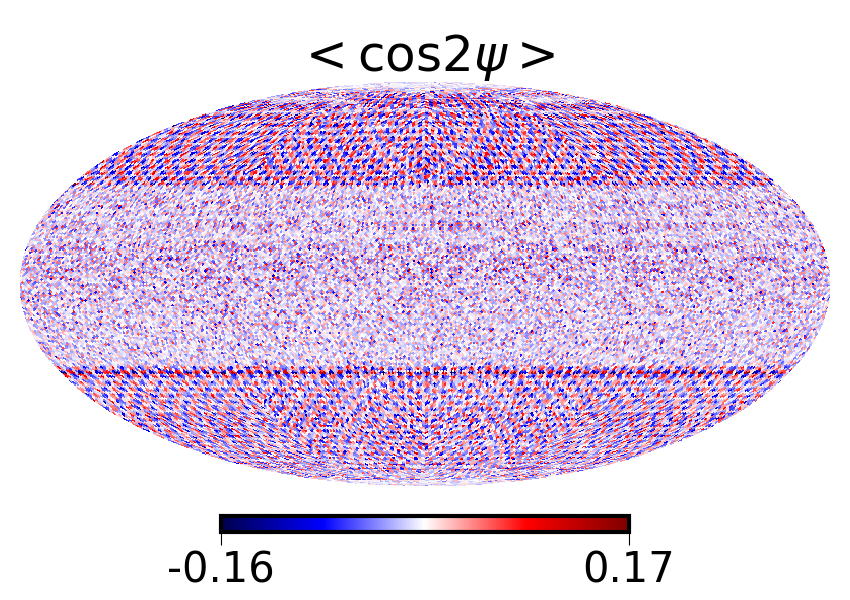}
\includegraphics[width=0.35\textwidth ]{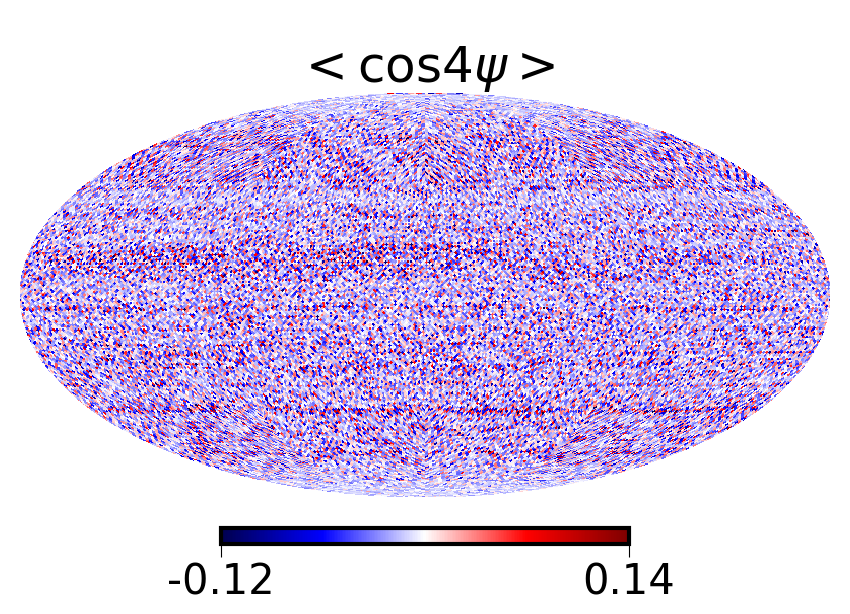}\\
\includegraphics[width=0.35\textwidth ]{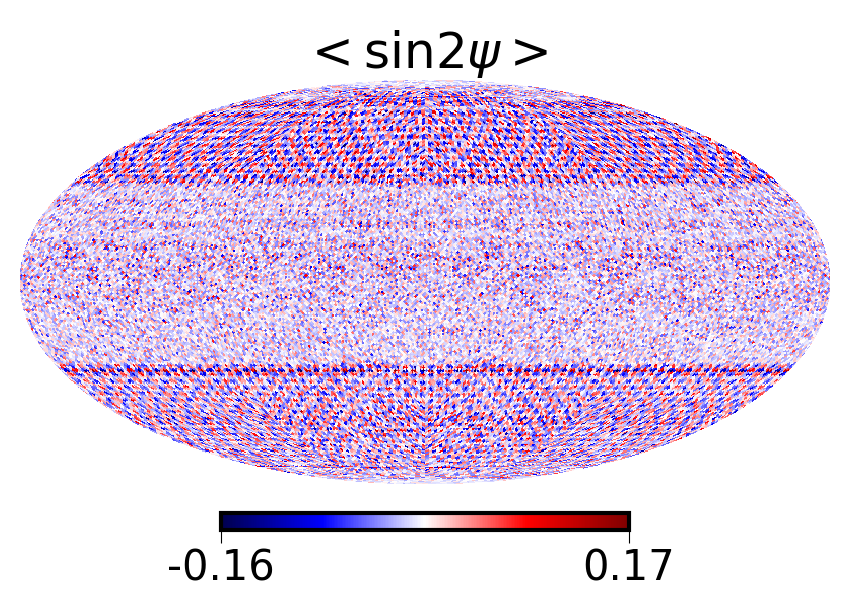}
\includegraphics[width=0.35\textwidth ]{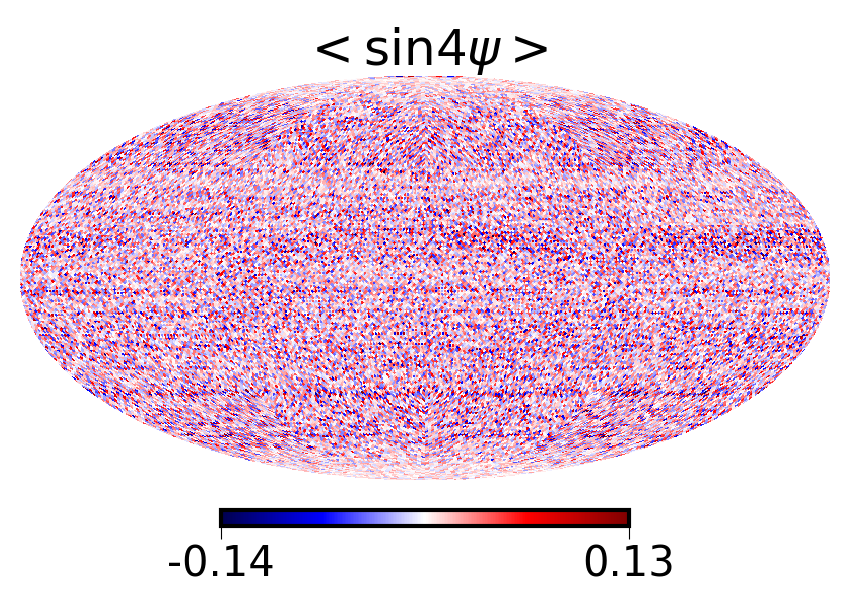}
\end{center}
\caption{{\bf Crossing moment maps (with rotating HWP).} 
We show the same maps as in Fig.~\ref{Fig:CrossLinkingMaps}
except that there is a rotating HWP, as explained in 
the main text. We observe that the coherent power in the cosine
maps has been scrambled as a result of the presence of the HWP
and the overall power in all the maps has greatly been reduced.
}
\label{Fig:CrossMapsHWP}
\end{figure}
\begin{figure}[hb]
\begin{center}
\includegraphics[width=0.38\textwidth ]{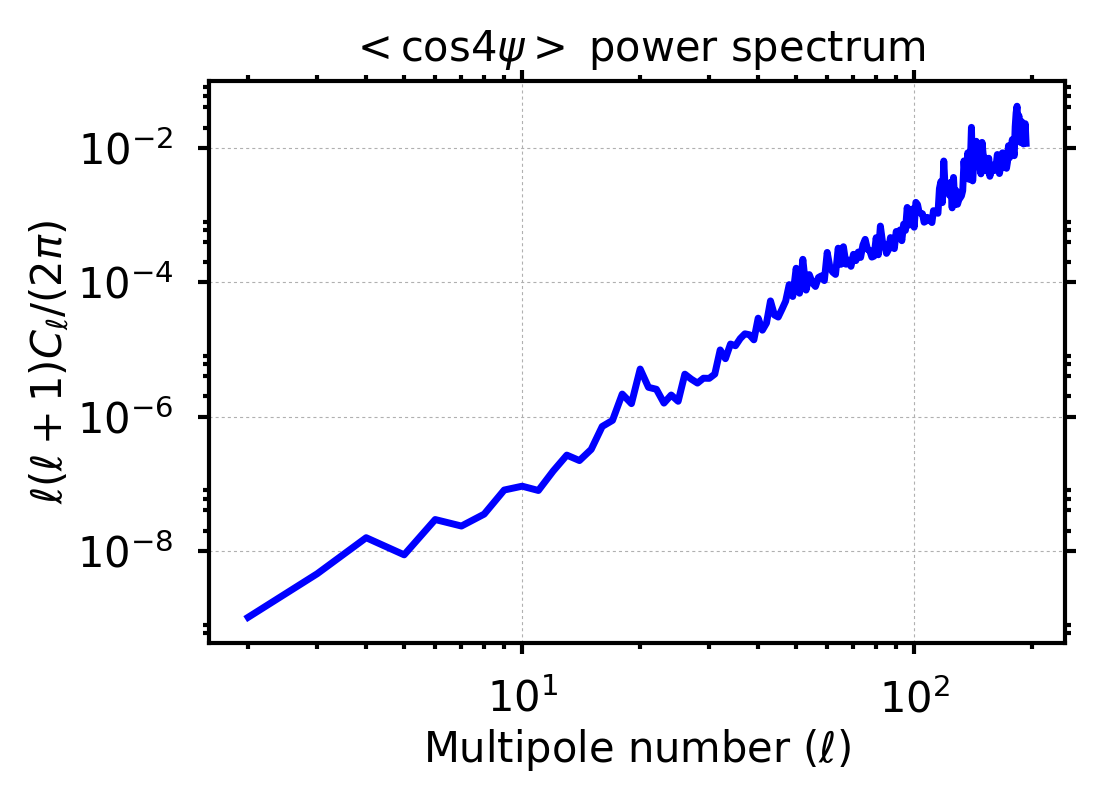}
\includegraphics[width=0.38\textwidth ]{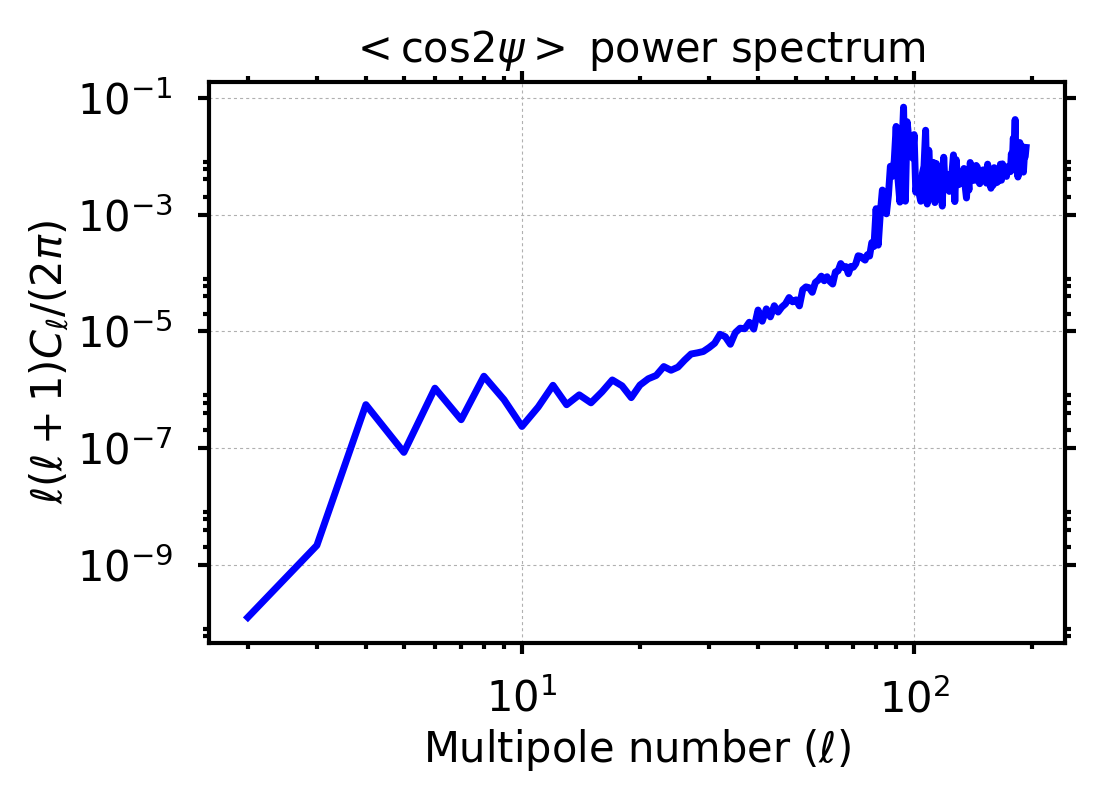}
\includegraphics[width=0.38\textwidth ]{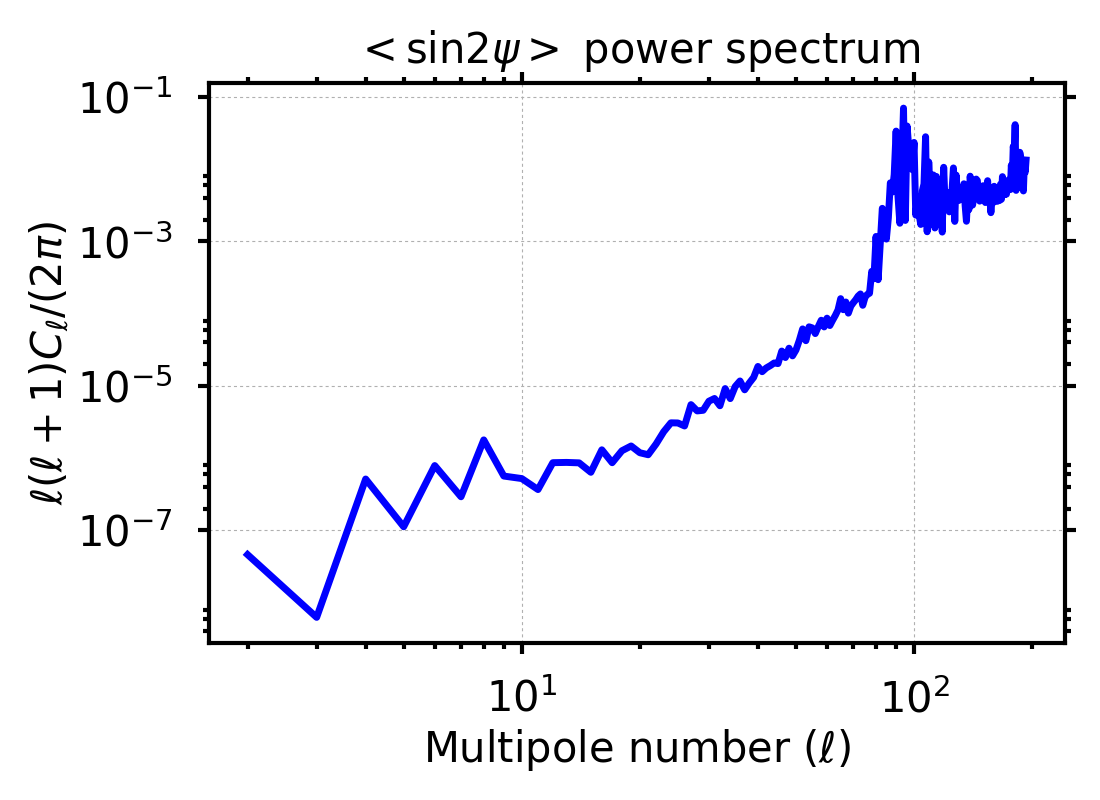}
\includegraphics[width=0.38\textwidth ]{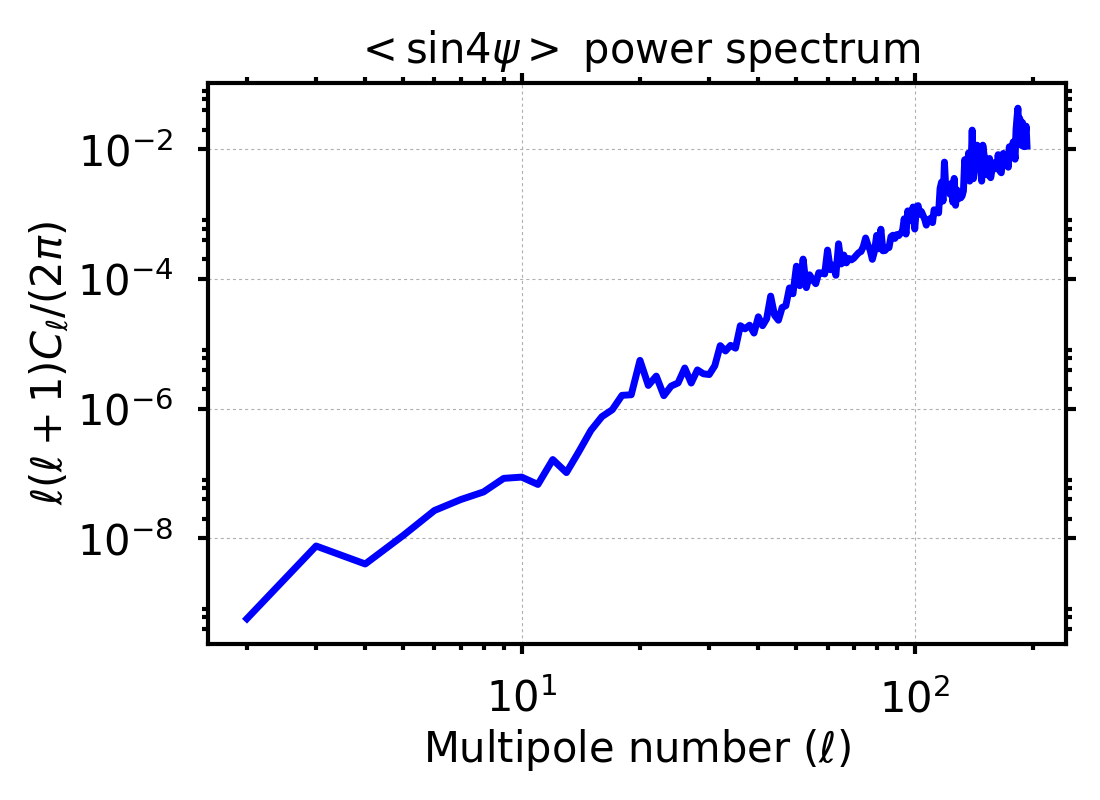}
\end{center}
\caption{{\bf Crossing moment map power spectra (with rotating HWP)} 
We show the power spectra for the maps in Fig.~\ref{Fig:CrossMapsHWP}.
The power spectra of the $\left< \cos 2\psi \right> $ and $\left< \cos 4\psi \right> $ 
have a white noise-like spectrum rather than an $(\ell ^{-2})$-like spectrum because
the HWP has scrambled on large-scale coherent azimuthally symmetry present in the case
with no HWP.
}
\label{Fig:CrossHWP:PS}
\end{figure}

\subsection{Hitcount and crossing moment map properties}

We now examine the properties of the hitcount map $H_a(p)$ for a particular detector
labeled by the index $a$ (where the index $p$ denotes a particular
discrete pixel) as well as maps of $\left< \cos 2\psi (p)\right> _a,$
$\left< \sin 2\psi (p)\right> _a,$ $\left< \cos 4\psi (p)\right> _a,$
and $\left< \sin 4\psi (p)\right> _a,$ which, as already stressed enter into the 
expressions for the bandpass mismatch.

Figures \ref{Fig:HitcountMaps} and \ref{Fig:CrossLinkingMaps}
show the maps $H_a(p),$ $\left< \cos 2\psi (p)\right>
_a,$ $\left< \sin 2\psi (p)\right> _a,$ $\left< \cos 4\psi (p)\right>
_a,$ and $\left< \sin 4\psi (p)\right> _a$ for a typical detector
with the fiducial scan parameters given above for a
full-year scan (so that there are no boundaries).

These figures demonstrate that in all the maps (except for the
$\left< \sin 2\psi (p)\right> _a,$ and $\left< \sin 4\psi (p)\right>_a$ maps), when 
small-scale structure is ignored, there is an azimuthally symmetric non-uniformity. 
From the azimuthally averaged quantities shown in
Fig.~\ref{Fig:CrossLinkingAzimuthalAverages}, we can see that superimposed on this 
azimuthally symmetric component is a component almost completely devoid of large-angle 
power resulting from the discreteness of the scans. Fig.~\ref{Fig:CrossLinkingPS} shows 
the power spectra of the crossing moment maps.
We note that given the finite size of the focal
plane, the spin opening angle $\beta $ varies from detector to
detector. This variation in $\beta $ induces an azimuthally symmetric
component having large-scale power in the difference map of moments for different detectors at different locations in the focal plane. Also
present will be a small-scale component, which would disappear in the
limit $\omega _{\rm spin},$ $\omega _{\rm prec}\to +\infty $ along with the
sampling rate while keeping the ratio $\omega _{\rm spin}/\omega _{\rm prec}$
fixed. This small scale power is somewhat akin to shot noise. 

We now consider the effect of a continuously rotating HWP on the second- and fourth-
order crossing moment maps, as shown in Figs.~\ref{Fig:CrossMapsHWP} and 
\ref{Fig:CrossHWP:PS}. We see that the azimuthally symmetric structures coherent on 
large angular scales disappear as a consequence of the continuously rotating HWP. The 
main consequence is to beat down by many orders of magnitude the $(\ell ^{-2.5})$-like 
power present on large angular scales in cosine maps, but there is also substantial
reduction in the power at all multipole numbers compared to the no-HWP case. 

We point out that much the same beneficial effect could also be obtained using a 
discretely stepped HWP (with a stepping pattern tailored to produce the necessary 
cancellations). Alternatively, less complete cancellations could also be obtained by 
stepping the orientation of the focal plane about its optical axis. These rotations are 
called ``deck rotations'' in the BICEP2 papers (see e.g.,~\cite{Bicep2KEK15}), a 
terminology that we shall also adopt. Allowing for such deck rotations, however, would 
also require additional complexity in the satellite design beyond the simplest no HWP 
design. Moreover, for the deck rotations alone, the cancellations would be imperfect 
because the values of $\beta $ for the individual detector scanning patterns change as 
the focal plane is rotated (except possibly for one detector situated at the optical 
axis, assumed to coincide with the deck rotation axis).

\clearpage
\newpage

\section{Conclusions}

This paper presented estimates of the contribution of bandpass mismatch error to the 
final determination of the tensor-to-scalar ratio $r,$ both for the window situated at
the `re-ionization bump,' and for the window at the `recombination bump', for a set of 
observation strategies considered for future CMB polarization experiments. In the case 
without a HWP, requiring in the optimal case the combination of
multi--detector data, we show that the bandpass mismatch error in polarization has 
a red power spectrum resembling $\ell ^{-2.5}.$ The contribution to $r$ is of the 
order of $10^{-3}$ at the reionization bump, assuming random variations of the detector 
filters for typical arrays at 140\,GHz, such that the variation of the dust component 
amplitude is of the order of 0.6\,\%.
However, with a continuously rotating HWP the spectrum is similar to that of
white noise, with the power on the largest scales many orders of magnitude smaller than 
without a HWP. This is due to the fact that an ideal HWP allows nearly 
uniform angle coverage in each pixel, and hence the multi-detector solution is almost 
equivalent to the combination of single detector maps of $Q$ and $U$. The HWP also 
cancels correlations in the non-uniformity in the angular coverage between 
different pixels, hence the efficient reduction in power of the bandpass mismatch 
on large angular scales. We further note that a stepped HWP would reduce bandpass mismatch in a similar way 
provided that its discrete rotations are properly synchronized with the scan pattern.
We show that even with a simplistic multi-detector
map-making approach, the HWP suppresses the bandpass leakage 
power by several orders of magnitude on large scales. We note however
that this conclusion ignores the problem of 
HWP imperfections, in particular chromaticity effects, which would 
generate bandpass mismatch systematics of its own.

To obtain accurate estimates of the bandpass mismatch
error, more precise information would be needed concerning (1)
the scan pattern assumed, (2) the variations in the bandpass
functions from detector to detector, and/or as the HWP rotates in case such a 
modulation is implemented, and (3) the foreground removal process. For
(1) we used one of the LiteBIRD candidate scan patterns.
Likewise, for (2) we based our model for variations in the 
bandpass function from preliminary results that have actually been achieved in the 
laboratory between different detectors without a HWP, but there may be 
effects not properly taken into account that could lead to
larger errors, or conversely further technological development
could lead to reduced mismatch between bandpass functions. 
With respect to (3), we simply calculated the bandpass error
in a 140\,GHz map, assuming that but for this error, the dominant
dust and synchrotron components could be removed by subtraction
using a perfect foreground component templates. 
This is certainly a simplification which provides a simple estimate that can be described in a simple
term. If the foregrounds turn out to be very complicated, the CMB
clean map might be the result of a linear combination of maps whose
coefficients (or varying sign) are much larger than one (assuming
the maps are normalized to the CMB). A foreground cleaning of this
sort (if necessary) may lead to larger bandpass errors than our estimate. 
Finally, we mention one caveat of our analysis: we did not include $1/f$ noise
in our modeling, a feature that allowed us to carry out pixel-by-pixel
map making and avoid including extra model parameters.%% Although we do not 
%%believe that including $1/f$ noise 
%%would substantially change our conclusions, we have not yet actually 
%%proven this.

In this paper we have estimated bandpass mismatch error assuming that
no measures have been taken to correct for or otherwise mitigate this
systematic error. In the companion paper Ref.~\cite{BPMMII} we explore paths to correct 
for and mitigate bandpass mismatch error with a dedicated data processing step.


\acknowledgments
Duc Thuong Hoang thanks the Vietnam International Education Cooperation Department (VIED) 
of the Ministry of Education and Training for support through a Ph.D.~fellowship 
grant. We thank Aritoki Suzuki for useful discussions and sharing with us data on filter bandpass measurements.

%Some of the results in this paper have been derived using the HEALPix package:\\ %http://healpix.sourceforge.net

\bibliographystyle{JHEP}
\bibliography{BiblioBandPass}

%\begin{thebibliography}{99}

% Please avoid comments such as "For a review'', "For some examples",
% "and references therein" or move them in the text. In general,
% please leave only references in the bibliography and move all
% accessory text in footnotes.

% Also, please have only one work for each \bibitem.

%\end{thebibliography}
\end{document}